\documentclass[11pt]{article}
\usepackage{amsfonts}
\usepackage{amssymb}
\def\bsh{\backslash}
\def\bdt{\dot \beta}
\def\adt{\dot \alpha}

\newfont{\bbbold}{msbm10 scaled \magstep1}

\def\bbC{\mbox{\bbbold C}}

\def\cA{{\cal A}}

\newfont{\goth}{eufm10 scaled \magstep1}

\def\gl{\mbox{\goth l}}

\def\gs{\mbox{\goth s}}

\def\a{\alpha}
\def\b{\beta}
\def\c{\gamma}
\def\d{\delta}\def\D{\Delta}
\def\e{\epsilon}
\def\f{\phi}
\def\vf{\varphi}

\def\l{\lambda}\def\L{\Lambda}
\def\m{\mu}

\def\p{\pi}
\def\P{\Pi}

\def\th{\theta}

\def\be{\begin{equation}}\def\ee{\end{equation}}
\def\bea{\begin{eqnarray}}\def\eea{\end{eqnarray}}
\def\ba{\begin{array}}\def\ea{\end{array}}

\def\del{\partial}

\def\str{\rm str}
\def\xz{\times}

\def\del{\partial}
\def\sdet{{\rm sdet}}
\def\str{{\rm str}}
\let\la=\label

\let\bm=\bibitem

\def\nn{\nonumber}
\def\bd{\begin{document}}
\def\ed{\end{document}}
\def\bea{\begin{eqnarray}}
\def\eea{\end{eqnarray}}
\def\ba{\begin{array}}
\def\ea{\end{array}}
\def\ft#1#2{{\textstyle{{\scriptstyle #1}\over {\scriptstyle #2}}}}
\def\fft#1#2{{#1 \over #2}}
\newcommand{\eq}[1]{(\ref{#1})}
\def\eqs#1#2{(\ref{#1}-\ref{#2})}
\def\det{{\rm det\,}}
\def\tr{{\rm tr}}
\newcommand{\ho}[1]{$\, ^{#1}$}
\newcommand{\hoch}[1]{$\, ^{#1}$}
\newcommand{\tamphys}{\it\small Center for Theoretical Physics,
Texas A\&M University, College Station, TX 77843, USA} 
\newcommand{\newton}{\it\small Isaac Newton Institute for Mathematical
Sciences, Cambridge, UK} 
\newcommand{\kings}{\it\small Department of Mathematics, King's College,
London, UK} 
\newcommand{\lapp}{\it\small LAPP, Annecy, France}
\newcommand{\HA}{{\Bbb H}\hskip-4.1pt {\Bbb A}}
\newcommand{\HR}{{\Bbb H}\hskip-4.7pt {\Bbb R}}
\newcommand{\dslash}{\partial\hskip-6.28pt /} 
\begin{document}
 \thispagestyle{empty}
 
 \hfill{KCL-MTH-99-46}

 \hfill{LAPTH-778/2000}

 \hfill{LTH-466}

 \hfill{hep-th/0001138}

 \hfill{\today}
\vspace{20pt} 

 \begin{center}
{\Large{\bf Four-point functions in $N=2$ superconformal field 
theories}} \vspace{30pt}\\ {\large B. Eden\hoch{a}, P.S. 
Howe\hoch{b}, A. Pickering\hoch{c}, E. Sokatchev\hoch{a} and P.C. 
West\hoch{b}} \vspace{15pt} 
\begin{itemize}
 \item [$^b$] \kings
 \item [$^a$] {\it\small Laboratoire d'Annecy-le-Vieux de Physique
 Th{\'e}orique\footnote{UMR 5108 associ{\'e}e {\`a}
 l'Universit{\'e} de Savoie} LAPTH, Chemin de Bellevue, B.P. 110,
F-74941 Annecy-le-Vieux, France} 
\item [$^c$] {\it\small Division of Theoretical Physics, Department of Mathematical Sciences, University of Liverpool, UK}  
 \end{itemize}
\vspace{60pt} 
 {\bf Abstract}
\end{center}
Four-point correlation functions of hypermultiplet bilinear 
composites are analysed in $N=2$ superconformal field theory using 
the superconformal Ward identities and the analyticity properties 
of the composite operator superfields. It is shown that the 
complete amplitude is determined by a single arbitrary function of 
the two conformal cross-ratios of the space-time variables.  

{\vfill\leftline{}\vfill \baselineskip=12pt \pagebreak 
\setcounter{page}{1} 

\section{Introduction}

In this paper we examine in detail the constraints imposed by 
superconformal invariance and analyticity on four-point 
correlation functions of analytic operators in four-dimensional 
superconformal field theories with $N=2$ supersymmetry. The 
analytic operators under consideration are gauge-invariant 
products of hypermultiplets which are represented by analytic 
superfields in $N=2$ harmonic superspace. Analytic superfields 
obey a generalised chirality-type constraint and depend 
holomorphically on the coordinates of the two-sphere which is 
adjoined to Minkowski superspace to form harmonic superspace. The 
definition of analyticity is given in more detail below. 

The sphere can be thought of as the homogeneous space $U(1)\bsh 
SU(2)$ and fields carry a charge with respect to the $U(1)$ 
isotropy group of this internal space. We shall be particularly 
interested in the case where each of the operators has charge 2. 
These operators are hypermultiplet bilinears and also have 
dimension 2. For the particular case of $N=4$ super Yang-Mills 
theory (SYM) there is a single hypermultiplet transforming under 
the adjoint representation of the gauge group and from this and 
its conjugate one can construct three charge 2 analytic bilinears. 
They can be viewed as $N=2$ components of the $N=4$ supercurrent, 
so that the corresponding four-point correlation functions give 
$N=2$ projections of the $N=4$ four-point correlator of four 
supercurrents. This correlator, or particular spacetime components 
of it, has been much studied in the context of the Maldacena 
conjecture \cite{maldacena} on both the AdS \cite{ads} and field 
theory sides \cite{4pt,4pt'}. 

The study of correlation functions of the above type in the 
harmonic superspace setting has been advocated in a series of 
papers \cite{hw1,hw2} \footnote{For an analysis of superconformal 
field theories in Minkowski superspace see, for example, 
\cite{op}}. The exact functional forms for two- and three-point 
functions were given \cite{hw1,hsw} (see also 
\cite{3pt,dhfs,dhfs2}) and it was later argued that the 
coefficients of such correlators in $N=4$ should be 
non-renormalised \cite {ehw} using analytic superspace techniques, 
the reduction formula and the notion of $U(1)_Y$ symmetry 
introduced in \cite{ken1}. (For the supercurrent correlators this 
non-renormalisation can also be seen as a consequence of anomaly 
considerations \cite{dhfs,hsw}.) It was also conjectured that 
four-point correlators of operators with sufficiently low charges 
might be soluble \cite{hw1,hw2}. However, in spite of the claims 
made for future work in these references, this turns out not to be 
the case and one purpose of the present work is to give the 
precise result that one finds for four charge 2 operators, which 
are known to be non-trivial \cite{4pt,4pt'}. As we shall show, the 
requirement of analyticity leads to additional constraints beyond 
those that one might expect on grounds of superconformal symmetry 
alone, but these are not enough to determine the correlation 
function completely. Elsewhere the result for charge 2 operators 
has been used to simplify the computation of the four-point 
function at two loops in perturbation theory \cite{ehssw2}. 

The work described here has been carried out over a period of 
several years and has focused on four-point functions of operators 
with equal charges. As we have just mentioned, these results are 
not as strong as had been conjectured. More recently, however, it 
has become apparent that more striking results can be obtained by 
considering asymmetric sets of charges. In \cite{ehssw3} the 
methods of the present paper were used to show that four-point 
extremal correlators ($p_4=p_1+p_2+p_3,\ p_i=$ charges) are simply 
given by products of free two-point functions. A discussion in $N=1$
perturbation theory is given in \cite{biakov}. Since the analogous
result had already been found in AdS supergravity \cite{dfmmrext}
(see also \cite{arufrov}), this also established a part of the Maldacena
conjecture. So, although the initial conjectures of \cite{hw1,hw2} 
for equal charges have turned out to be incorrect, perhaps it is 
not unreasonable to claim some partial vindication for the initial 
optimism in the light of the fact that the very same strategy 
gives a good way of proving the recent results for extremal 
correlators in a field-theoretic setting.

The analysis of analytic operators and correlation functions can 
be carried out in two equivalent frameworks each having their own 
distinctive features. One can either work in analytic superspace 
(with complexified spacetime) using explicit coordinates, or one 
can work in harmonic superspace (with real spacetime) using an 
equivariant formalism with respect to the internal $SU(2)$ 
symmetry group. In the former approach, all the coordinates appear 
on an equal footing. This makes the action of the superconformal 
group is more transparent and facilitates the construction of 
invariants. Further, analyticity (or the lack of it) is manifest. 
In the latter approach the internal $SU(2)$ is treated 
covariantly, so that explicit coordinates for the sphere do not 
have to be introduced. In addition, harmonic analyticity can be 
interpreted as an irreducibility condition under $SU(2)$. This 
allows one to limit the analysis to the first two non-trivial 
levels in the $\theta$ expansion of the correlator. We shall 
employ both methods in this paper, using the analytic formalism in 
section 2 and the covariant formulation in section 3.


\section{Coordinate approach \label{CoordForm}}

\subsection{Analytic superspace}

$N=2$ harmonic superspace, first introduced in \cite{gikos}, is 
the product of $N=2$ Minkowski superspace and the two-sphere 
$S^2=\bbC P^1$. A field on this space can be expanded in harmonics 
on the sphere with coefficients which are ordinary $N=2$ 
superfields. A general superfield on $N=2$ harmonic superspace is 
therefore equivalent to an infinite number of ordinary 
superfields, but, if a superfield depends holomorphically on the 
coordinates of the sphere, it will have a finite expansion in 
ordinary superfields due to the fact that spaces of holomorphic 
tensors on the sphere are finite-dimensional. Such a field is 
called harmonic analytic (H-analytic). In addition one can define 
a generalised notion of chirality, called Grassmann analyticity 
(G-analyticity) such that a G-analytic field depends on only half 
the number of odd coordinates of the full superspace. A field 
which is both H-analytic and G-analytic will be called analytic. 

An alternative description of analytic superfields on complexified 
Minkowski superspace is in terms of (holomorphic) fields on 
analytic superspace, a space with half the odd dimensionality of 
harmonic superspace. It bears a similar relation to harmonic 
superspace as chiral superspace does to Minkowski superspace. 

Analytic superspace is a homogeneous space of the complexified 
$N=2$ superconformal group, $SL(4|2)$ (for a review of various 
homogeneous superspaces in this context see \cite{hh}). This group 
acts naturally (to the left) on $N=2$ supertwistor space 
$\bbC^{4|2}$, and the coset we are interested in has the 
geometrical interpretation as the Grassmannian of $(2|1)$-planes 
in twistor space. The body of the whole of this Grassmannian is 
compact, whereas we are interested only in the usual, non-compact 
Minkowski spacetime. The body of the actual space we shall work 
with will therefore be restricted in this sense, although its 
internal part, $\bbC P^1$, remains compact. This is similar in 
spirit to regarding $\bbC$ as an open subset of $\bbC P^1$ 
obtained by omitting the point at infinity. We shall be a bit more 
precise about this below, after we have introduced appropriate 
local coordinates. 

In the usual basis for $N=2$ supertwistor space the first four 
elements correspond to the even  part and the second two to the 
odd part. If we instead make a choice of basis ordered in the 
sequence two even, one odd, two even, one odd, the isotropy group, 
$H$, of analytic superspace will consist of supermatrices of the  
simple form 

\be
\left( \ba{cc} \bullet & 0\\ \bullet &\bullet \ea \right) \; . \ee 

Here each entry represents a $(2|1)\xz (2|1)$ supermatrix and the 
bullets denote non-singular such matrices. The superdeterminant is 
constrained to be 1. In this basis it is reasonably clear that the 
coset space $H\bsh SL(4|2)$ is indeed the Grassmannian of 
$(2|1)$-planes in $\bbC^{4|2}$. 

Using standard homogenous space techniques we may choose a local 
coset representative $s(X), X\in M_A$ (analytic superspace) as 
follows: 

\be
M_A\ni X\rightarrow s(X)=\left( \ba{cc} 1 & X\\ 0 &1 \ea 
\right)\in SL(4|2) \ee 

and again each entry represents a $(2|1)\xz (2|1)$ supermatrix. 
The components of $X$ are given by 

\be
X=\left( \ba{cc} x^{\a\adt}& \l^{\a}\\ \p^{\adt} &y\ea \right) \ee 

where $\a,\adt$ are two-component spinor indices, $x$ is the 
spacetime coordinate, $\l$ and $\p$ are the odd coordinates and 
$y$ is the standard coordinate on $\bbC P^1$. (We shall 
occasionally use index notation $X^{AA'}$ for the coordinate 
matrix $X$.)  As we mentioned above, we want the body of our 
superspace to consist of non-compact spacetime together with a 
compact internal space, $\bbC P^1$. The full space we are 
interested in can be covered by two open sets corresponding to the 
two standard open sets of the sphere. If we denote these two sets 
by $U$ and $U'$, and put primes on the coordinates for $U'$ we 
find that the two sets are related as follows on the overlap: 

\bea x'&=& x-{\l\p\over y} \; , \nn\\ \l'&=& {1\over y}\l \; , \nn 
\\ \p'&=& {1\over y}\p \; , \nn\\ y'&=&{1\over y} \; . \eea 

We see that the odd coordinates and the coordinates of the 
internal space together parametrise $\bbC^{(1|4)}$, so that the 
whole space has the form of an affine bundle of rank $(4|0)$ over 
$\bbC P^{(1|4)}$. 

Superconformal transformations can be discussed straightforwardly 
using standard homogeneous space methods. Under a superconformal 
transformation $X\rightarrow X\cdot g,\ g\in SL(4|2)$. The 
transformed coordinates $X\cdot g$ are determined using the 
formula 

\be
s(X\cdot g)=h(X,g)s(X)g \; . \ee 

The r\^{o}le of the compensating transform $h(X,g)$ (an element of 
$H$) is to restore the form of the coset representative $s(X)$ 
after right-multiplication by $g$. For an infinitesimal 
transformation specified by $\cA\in \gs\gl (4|2)$, the Lie 
superalgebra of $SL(4|2)$, we have 

\be
\d X=B + AX + X D + XCX \la{sct} \ee 

with 

\be
\cA=\left( \ba{cc} -A& B\\ -C &D\ea \right) \, . \ee 

An analytic field of charge $p$ is a field $\f$ on analytic 
superspace which transforms under an infinitesimal superconformal 
transformation according to the rule 

\be
\d\f =V\f + p\D\f \ee 

where  $V$ is the vector field generating the transformation, 
$V=\d X{\del\over\del X}$, and $\D:=\str(A+XC)$. The free, 
on-shell hypermultiplet is such a field with charge $1$ and will 
be discussed in more detail below. In an interacting theory, the 
on-shell hypermultiplet is covariantly analytic, but 
gauge-invariant products of hypermultiplets are analytic in the 
above sense with charges equal to the number of hypermultiplets in 
the product. 

For completeness we reproduce here the explicit expressions for 
the vector fields corresponding to the different types of 
superconformal transformation. {}From \eq{sct} one can read off 
the vector fields for each of the parameters. They divide into 
translational ($B$), linear ($A,D$) and quadratic ($C$) types. The 
translations are ordinary spacetime translations, half of the 
$Q$-supersymmetry transformations and translations in the internal 
$y$ space, $\bbC P^1$. The corresponding vector fields are 

\begin{equation}
V_{AA'}= {\del\over\del X^{AA'}} \label{trans} \; . 
\end{equation}

The linearly realised symmetries are Lorentz transformations 
($SL(2)\xz SL(2)$) in complex spacetime) and dilations, internal 
dilations, $R$-symmetry transformations, the other half of the 
$Q$-supersymmetries and half of the $S$-supersymmetries. The 
Lorentz transformations are handled in the usual way so that we do 
not need to write them down. The vector fields generating 
dilations ($D$), internal dilations ($D'$) and $R$-symmetry 
transformations are 

\begin{eqnarray} V(D) &=& x^{\a\adt}\del_{\a\adt} + {1\over2}(\l^{\a
'}\del_{\a}+\p^{\adt}\del_{\adt})\;, \\ V(D') &=& y\del_y + 
{1\over2}(\l^{\a }\del_{\a}+\p^{\adt}\del_{\adt})\;, \label{dil} 
\\ V(R)&=& \l^{\a}\del_{\a}-\p^{\adt}\del_{\adt} \;. 
\end{eqnarray} 

The vector fields generating linearly realised $Q$-supersymmetry 
are 

\begin{eqnarray} V(Q)_{\a} &=& \p^{\adt}\del_{\a\adt} +y\del_{\a}\;,\\ V(Q)_{\adt} &=& \l^{\a}\del_{\a\adt}-y\del_{\adt}\;, \label{linq} \end{eqnarray}

while those generating linearly realised $S$-supersymmetry are 

\begin{eqnarray} V(S)^{\a} &=& x^{\a\adt}\del_{\adt} + \l^{\a}\del_{y}\;,\\ \label{lins1} V(S)^{\adt} &=& x^{\a\adt}\del_{\a}-\p^{\adt}\del_{y}\;. \label{lins2} \end{eqnarray}

The remaining supersymmetry transformations are the non-linearly 
realised $S$-supersymmetries generated by 

\begin{eqnarray} V(S)^{\adt}&=& x^{\b\adt}\p^{\bdt}\del_{\b\bdt} +
x^{\b\adt}y\del_{\b} -\p^{\adt}\p^{\bdt}\del_{\bdt} -\p^{\adt} 
y\del_{y}\;,\\ V(S)^{\a}&=& -\l^{\b}x^{\a\bdt}\del_{\b\bdt} 
-\l^{\b}\l^{\a}\del_{\b} + yx^{\a\bdt}\del_{\bdt} + 
y\l^{\a}\del_{y}\;. \label{quads} \end{eqnarray} 

Finally, we have conformal boosts ($K$) and internal conformal 
boosts ($K'$) generated by 

\begin{eqnarray} V(K)^{\a\adt} &=& x^{\b\adt} x^{\a\bdt}\del_{\b\bdt}+
x^{\b\adt} 
\l^{\a}\del_{\b}+\p^{\adt}x^{\a\bdt}\del_{\bdt}+\p^{\adt}\l^{\a 
}\del_{y}\;,\\ V(K') &=& \l^{\b}\p^{\bdt}\del_{\b\bdt} + 
\l^{\b}y\del_{\b} + y\p^{\bdt}\del_{\bdt} + y^2\del_{y}\;. 
\label{boost} \end{eqnarray} 

The function $\D$ is non-zero in the following cases: 

\bea \D(D) \; \, &=& 1 \; , \nn\\ \D(D') \, &=& -{1\over2}  \; , 
\nn\\ \D(S)^{\adt}&=& \p^{\adt} \; , \nn\\ \D(S)^{\a}&=&-\l^{\a} 
\; , \nn\\ \D(K)^{\a\adt}&=&  x^{\a\adt} \; , \nn\\ \D(K') \, &=& 
-y \; . \eea 

In these lists for $V$ and $\D$ we have omitted the parameters 
which of course have the opposite indices to those displayed. The 
parameters can easily be restored, but this should be done from 
the left in order to get the right signs. 


\subsection{Analytic fields}

We recall that a holomorphic tensor field of charge $p$ on $\bbC 
P^1$ is given by two local functions $a,a'$ on $U,U'$ 
respectively, such that, in the overlap 

\be
a(y)=y^p a'(y') \ee 

Expanding both sides in power series in their respective 
variables, equating powers of $y(={1\over y'})$ and demanding the 
absence of poles we find 

\be
a(y)=\sum_{n=0}^{n=p} a_n y^n \; . \ee 

and similarly for $a'(y')$. The two expansions are related by 

\be
a_n = a'_{p-n} \; . \ee 

Hence the space of tensor fields of charge $p$ is a 
finite-dimensional space with dimension $p+1$ which can be 
identified with the space of $p$th rank totally symmetric tensors 
under $SL(2)$. 

In a similar fashion we define an analytic superfield of charge 
$p$ on $M_A$ to be specified by two local holomorphic functions 
$\f(x,\l,\p,y)$ and $\f'(x',\l',\p',y')$ defined on the two 
standard coordinate patches $U,U'$ respectively, such that, in the 
overlap 

\be
\f(x,\l,\p,y)=y^p \f'(x',\l',\p',y') \; . \ee 

If we now expand both sides in the odd variables and in $y$ or 
$y'$, we obtain restrictions on the component functions. For 
example, the zeroth order term in $\l\p$ is an $x$-dependent 
charge $p$ holomorphic tensor on $\bbC P^1$, the component of $\l$ 
behaves like a tensor of charge $p-1$ and so on. In addition, 
since the relation between $x$ and $x'$ involves a shift, 
spacetime derivatives will appear in the constraints. 

The basic superfield we shall consider is the hypermultiplet. In 
this language this multiplet is represented by an analytic 
superfield $\f$ of charge 1. Using the above method one can easily 
seen that it has only a short expansion: 

\be
\f(x,\l,\p,y)=\vf(x,y)+\l^{\a}\psi_{\a}(x) + \p^{\adt} 
\chi_{\adt}(x) +\l^{\a}\p^{\adt} \check \vf_{\a\adt}(x) \ee 

Furthermore 

\be
\vf(x,y)=\vf_o(x) + y \vf_1(x) \ee 

and similarly for $\vf'(x',y')$, with 

\be
\vf_o=\vf'_1;\qquad \vf_1=\vf'_o \; . \ee 

We also find 

\be
\check\vf_{\a\adt}=-\del_{\a\adt} \vf_1;\qquad 
\check\vf'_{\a\adt}=\del_{\a\adt}\vf_o \; . \ee 

In addition, the fields $\vf_o,\vf_1,\psi,\chi$ must all satisfy 
their equations of motion 

\bea \square \vf_o=\square \vf_1 &=& 0 \; , \nn \\ 
\del^{\adt\a}\psi_{\a} &=& 0 \; , \nn\\ \del^{\a\adt}\chi_{\adt} 
&=&0 \; . \eea 

This is the usual hypermultiplet with two complex scalar fields 
and two complex Weyl fermions, all of which are physical and 
on-shell. In the interacting theory, the on-shell hypermultiplet 
will be covariantly analytic, but gauge-invariant products of 
hypermultiplets will be analytic tensor fields of the type we have 
just described with charges $2,3,\ldots$. For example, a field of 
charge two contains an independent vector field $v_{\a\adt}$, 
which is conserved, but there are no equations of motion. In other 
words a charge two field is a linear multiplet. 


\subsection{Correlation functions and Ward identities}

In this section we consider the superconformal Ward identities for 
four-point correlation functions for analytic operators with 
charges $p_1,p_2, p_3, p_4$. We shall denote such correlators by 
$<p_1 p_2 p_3 p_4>$. Note, however, that the operators are not 
assumed to be the same even if they have the same charges, in 
particular, we shall not impose any symmetry requirements on such 
correlators. 

If we assume that analyticity holds in the quantum theory, the 
superconformal Ward identity for such a correlator reads 

\be
\sum_{i=1}^4 (V_i + p_i\D_i) <p_1 p_2 p_3 p_4>=0 \ee 

The assumption of analyticity is tantamount to using the field 
equations of the underlying hypermultiplet at operator level. This 
should be reasonable provided that we keep the points separated. 
In the case of charge two operators, which is the main focus of 
this paper, the H-analyticity condition implies, as we noted 
above, that the superfield describes a linear multiplet with a 
conserved spacetime current. Analyticity can be examined directly 
in perturbation theory using the harmonic superspace formalism and 
has been verified in all the examples that have been looked at so 
far. 

We shall consider correlators of the above type which have 
non-vanishing leading terms. Further, we will specialise to 
correlators with four equal charges $p$. In this case we can write 

\be
<pppp>=(g_{12})^p (g_{34})^p F \ee 

where $g_{ij}$ is the free two-point function for charge one 
operators at points $i$ and $j$, 

\be
g_{ij}=\sdet X_{ij}^{-1}={\hat y_{ij}\over x_{ij}^2} \ee 

and $F$ is an arbitrary function of invariants. Here 
$X_{ij}=X_i-X_j$ denotes the  coordinate difference matrix for 
points $i$ and $j$ and 

\be
\hat y_{ij}=y_{ij}-\p_{ij} x_{ij}^{-1} \l_{ij} \ee 

with the index convention that $x^{-1}$ has a pair of subscript 
indices $\adt\a$. 

It is straightforward to check that $g_{ij}$ satisfies the 
equation 

\be
(V_i+V_j +\D_i + \D_j) g_{ij}=0 \ee 

so that the Ward identity for $<pppp>$ will indeed be satisfied 
for any invariant $F$. 

It is not difficult to see that there must be more $N=2$ analytic 
superconformal invariants than there are spacetime conformal 
invariants for four points. If we expand $F$ in the odd variables 
$\l$ and $\p$ the leading term must be invariant under spacetime 
conformal transformations and also under conformal ($SL(2)$) 
transformations of $\bbC P^1$. At four points there are two 
independent spacetime ($x$) cross-ratios and one independent 
internal ($y$) cross-ratio, so that there should be at least three 
independent analytic superconformal invariants at four points. In 
fact, there are no more. Any additional independent invariant 
would have to be nilpotent, but one can easily show that there are 
no such invariants at four points. The reason is essentially due 
to counting; there are $4\xz 4=16$ odd coordinates which is 
precisely equal to the number of supersymmetries in $SL(4|2)$. On 
a putative nilpotent invariant these supersymmetries behave 
essentially like translational symmetr! ies so that any possible 
leading term must vanish. In more detail, suppose that $F$ is a 
nilpotent four-point invariant. It can be written $F_o + ...$ 
where $F_o$ is the term with the lowest power of $\l \p$. By 
examining the supersymmetry transformations directly one can see 
that the first term in the variation of $F$ involves only $F_o$. 
Furthermore, again by looking at each transformation in turn, one 
finds that setting this first term in the variation equal to zero 
(because $F$ is an invariant) leaves no possible solutions for 
separated points. Consequently one concludes that there can be no 
nilpotent invariants. This argument is presented in detail in 
\cite{ehw}. 

It was shown in \cite{hw2} that non-nilpotent analytic superspace 
superconformal invariants can be expressed in terms of 
superdeterminants and supertraces of the coordinate differences 
$X_{ij}$. For the case in hand a possible choice of three 
independent invariants is given as follows: we take two super 
cross-ratios 

\be
S={\sdet X_{14}\sdet X_{23}\over\sdet X_{12}\sdet X_{34}} \; , 
\qquad T={\sdet X_{13}\sdet X_{24}\over\sdet X_{12}\sdet X_{34}} 
\ee 

and one supertrace invariant 

\be
U=\str(X_{12}^{-1}X_{23} X_{34}^{-1} X_{41}) \; . \ee 

The invariants $S$ and $T$ may be expressed in terms of the 
spacetime cross-ratios 

\be
s={x_{14}^2 x_{23}^2\over x_{12}^2 x_{34}^2} \; , \qquad 
t={x_{13}^2 x_{24}^2\over x_{12}^2 x_{34}^2} \ee 

and the internal cross-ratio 

\be
v={y_{14}y_{23}\over y_{12} y_{34}} \ee 

in the form 

\be
S={s\over \hat v} \; , \qquad T={t\over \hat w} \; . \ee 

Here $w={y_{13}y_{24}\over y_{12} y_{34}}=1+v$, and the hats on 
$v$ and $w$ are defined by hatting each of the $y$ variables in 
their definitions, 

\be
\hat v={\hat y_{14}\hat y_{23}\over \hat y_{12} \hat y_{34}} \;, 
\qquad \hat w={\hat y_{13}\hat y_{24}\over \hat y_{12} \hat 
y_{34}} \;. \ee 

We may write 

\be
\hat w=1+\hat v+\D w \ee 

and express the third invariant $U$ in the form 

\be
U=1-t+s +\hat v +\D U \; . \ee 

Both $\D W$ and $\D U$ are nilpotent quantities. The explicit 
expressions are as follows: 

\bea \D U &=& {1\over \hat y_{12}\hat y_{34}} \Big(-\hat y_{12} 
\P_{2}x_{32}^1\L_2 + \hat y_{34} \P_1 x^4_{23}\L_1 \nn\\ 
&\phantom{=}&+\hat y_{23}\left( \P_{1}x_{21}^4\L_1 - 
\P_{2}x_{34}^1\L_2  - \P_{1}\bar x_{21}^4\L_2 +\P_{2}\bar 
x_{34}^1\L_1 \right)\hspace{2.5cm}\nn \\ 
&\phantom{=}&+(\P_1x_{23}\L_2)(\P_2 x_{23}\L_1) \Big) \la{u} \eea 

and 

\bea \D w &=& {1\over \hat y_{12}\hat y_{34}} \Big(-\hat y_{12} 
\P_{2}x_{32}^4\L_2 + \hat y_{34} \P_1 x^1_{23}\L_1 \nn\\ 
&\phantom{=}&+\hat y_{23}\left( \P_{1}[x_{23}^1-x_{24}^1]\L_1+ 
\P_{2}[x_{31}^4-x_{32}^4]\L_2  + \P_{1}\bar x_{24}^1\L_2 
-\P_{2}\bar x_{31}^4\L_1 \right) \nn\\ 
&\phantom{=}&-(\P_1x_{21}^3\L_1)(\P_2 x_{34}^2\L_2)\Big) \la{w} 
\eea 

where we have used the convenient shorthand 

\bea x_{ij}^k&=& x_{ik} x_{jk}^{-1} x_{ij}, \\ \bar x_{ij}^k&=& 
x_{ik} x_{jk}^{-1} x_{jl} \;\; l\ne, i,j,k \;. \eea 

The odd variables are defined by 

\bea 
\L_{1\adt}&=&(x_{12})^{-1}_{\adt\a}\l_{12}^{\a}-(x_{23})^{-1}_{\adt\a}\l_{23}^{\a}\;,\nn\\ 
\L_{2\adt}&=&(x_{23})^{-1}_{\adt\a}\l_{23}^{\a}-(x_{34})^{-1}_{\adt\a}\l_{34}^{\a} 
\eea 

and, similarly, 

\bea 
\P_{1\a}&=&\p_{12}^{\adt}(x_{12})^{-1}_{\adt\a}-\p_{23}^{\adt}(x_{23})^{-1}_{\adt\a}\;,\nn\\ 
\P_{2\a}&=&\p_{23}^{\adt}(x_{23})^{-1}_{\adt\a}-\p_{34}^{\adt}(x_{34})^{-1}_{\adt\a}\;. 
\eea 

Now the crucial point is the following: each of the operators in 
the correlator can be expanded as a polynomial in $y$, so that the 
correlator is manifestly analytic in the $y$ variables. On the 
other hand, each of the invariants depend on the $y$'s in a 
rational manner so that one might expect the absence of 
singularities to impose further constraints on $F$. These 
constraints will clearly depend on the charges involved. For the 
correlator $<pppp>$ the lowest term in $F$ must be of  the form 

\be
F|_{\l=\p=0}=a_0 + a_1 v + \ldots a_{p+1} v^{p+1} \ee 

where each of the $a$'s depends on the cross-ratios $s,t$. The 
question then arises whether there are further constraints on the 
coefficient functions at higher orders. It is clear that the lower 
the charge the more constrained $F$ must be. For charge one, there 
are  no gauge-invariant operators, so the simplest interesting 
case to examine is charge 2 to which we turn in the next section. 


\subsection {Analyticity analysis}

In this section we analyse the constraints imposed by 
H-analyticity on the four-point function of four charge two 
operators (not necessarily the same). It can be written 

\be
<2222>={\hat y_{12}^2\hat y_{34}^2\over x_{12}^4 x_{34}^4} 
F(S,T,U) \label{startingPoint} \ee 

The invariants $S,T,U$ are convenient from some points of view - 
it is easy to see that they are invariants, and they have concise 
explicit forms. Requiring H-analyticity implies that $F$ should 
have no singularities in 
$(\hat{y}_{13},\hat{y}_{14},\hat{y}_{23},\hat{y}_{24})$ and that 
it can have poles up to order two in $(y_{12},y_{34})$ (if we had 
charge $p$ operators this would become order $p$). Since $(\hat 
y_{12}\hat y_{34})$ occurs as a denominator in $S^{-1},T^{-1}$ and 
$U$ the regularity of the correlator will lead to constraints for 
$F$. 

Note that regularity in the hatted variables is equivalent to 
regularity in the unhatted ones: if we demand that the whole 
correlator be a polynomial in the hatted $y$'s and write, for 
example, $\hat y_{12}=y_{12}+\d_{12}$, then a Taylor expansion 
will produce a polynomial of the same degree in the unhatted 
$y$'s, because the $\d$'s are non-singular and $y$-independent. 

Clearly, H-analyticity should hold at each order in the odd 
variables separately. We shall carry out the expansion in two 
steps: we first expand in the nilpotent quantities $\D U,\D w$ and 
then express the latter in terms of products of spinors. We will 
here refer to the order in $\D U, \, , \D w$ as ``level'' in order 
to distinguish the first from the second step. The main technical 
problem one faces in this approach is to compute which of the 
various products of the form $(\D u)^p (\D w)^q$ are independent 
for each fixed value of $p+q$, i.e. at a given ``level''. 

From the point of view of expanding in odd variables it is 
helpful to think about an equivalent set of invariants, $S',T',V$, 
which are defined to have leading terms $s,t$ and $v$ 
respectively. These invariants are expressed in terms of $S,T,U$ 
by 

\be
S'=SV,\qquad T'=T(1+V),\qquad V={T+U-1\over 1+S-T} \ee 

Since we have 

\bea S'&=& s+\ldots \\ T'&=& t+\ldots \\ V&=& v+\ldots \eea 

it follows from (\ref{startingPoint}) that we may write $F$ in the 
form 

\be
F=a_1(S',T') + a_2(S',T')V + a_3(S',T') V^2 \; . \ee 

This expression clearly meets the lowest level requirements of 
analyticity for charge 2 and shows that the dependence on the 
third invariant $V$ is thereby fixed. The objective now is to 
Taylor expand F about $S'=s, T'=t$ and $V=\hat v$. However, since 
the original set of invariants is easier to evaluate explicitly, 
we shall convert the Taylor expansion back into these variables as 
we go. In this way we arrive at a power series in the nilpotent 
quantities $\D U$ and $\D w$. We will then express both of these 
and the coefficients that accompany them in terms of the set of 
coordinates $(x_{12},x_{23},x_{34},\hat y_{12},\hat y_{23}, \hat 
y_{34},\L_1,\L_2,\P_1,\P_2)$. These variables are convenient in 
the sense that they are invariant under translational 
$Q$-supersymmetry and linear $S$-supersymmetry, and in addition 
they involve no $y$-singularities. Furthermore, as noted above, 
the difference between $y$ and $\hat y$ is non-singular so it is 
permissible to study analyticity in the latter rather than the 
former. 

The Taylor expansion of $F$ is then 

\begin{eqnarray}
F(S,T,U) & = F_o + \Delta U (\partial_U F) + \Delta w (- 
\frac{t}{\bar{w}^2} \partial_T F) \nn\\ & \frac{1}{2} \Delta U^2 
(\partial_U^2 F) + \Delta U \Delta w (- \frac{t}{\bar{w}^2} 
\partial_U
\partial_T F) + \frac{1}{2} \Delta w^2 (\frac{t^2}{\bar{w}^4}
\partial_T^2 F + 2 \frac{t}{\bar{w}^3} \partial_T F) + ... 
\label{taylExp} 
\end{eqnarray}

where $F_o$ is $F(S',T',V)$ evaluated at $(s,t,\hat v)$. Clearly 

\be
F_o=a_1(s,t) + a_2(s,t) \hat v + a_3(s,t) \hat v^2 \;. 
\label{FForm} \ee 

In the new variables the partial derivatives, evaluated at 
$(S',T',\hat v)$, are 

\begin{eqnarray}
\partial_U &=& \frac{1}{R} \, D \; ,\nn \\
\partial_T &=& {(1 + \hat{v})\over R}(D + R\partial_t) 
\end{eqnarray}

with 

\bea R &=& s + \hat{v} - \frac{\hat{v} \, t}{1 + \hat{v}} 
\nonumber \;, \\ D &=& s \, \partial_s + \hat{v} \, 
\partial_{\hat{v}} + \frac{\hat{v} \, t}{1 + \hat{v}} \, 
\partial_t \; . \eea 

Here $v$ has been replaced by $\hat{v}$ because we wish to use the 
Taylor expansion about the point $(s,t,\hat{v})$. We will almost 
always multiply $R$ and $D$ by $\bar{w} = 1+\hat{v}$ in order to 
avoid the singularity in their denominators. Note that $DR = R$. 

We shall refer to the expressions in round brackets multiplying a 
certain power $\Delta U^p \Delta w^q$ in (\ref{taylExp}) as 
``component functions''. The expansion ends at fourth order, 
because $\Delta U, \Delta w$ are $R$-symmetric and are therefore 
power series in $(\Pi \Lambda)$. Since the spinors are 
two-component objects and since there are two $\P$'s and two 
$\L$'s it follows that the highest possible power is $(\Pi 
\Lambda)^4$. 

Let us investigate the linear level in the Taylor expansion. We 
ask whether singularities in $\Delta U$ and $\Delta W$ can 
conspire to cancel or whether the latter are independent objects. 
The component functions depend on $s,t,\hat{v}$, which yields a 
linear dependence problem with coefficients in the ring of 
functions of $s,t,\hat{v}$. 

The functional form of these coefficients can be made much more 
explicit. Given the form of $F$ (\ref{FForm}), by commuting all 
$R$'s to the left and all $\partial_t$'s to the right we can show 
that the general component functions at $k$-th level have the form 

\begin{equation}
\frac{1}{(\bar{w} (\bar{w} R))^k} \, \sum_{n=0}^{2 + 2 k} c_n(s,t) 
\, \hat{v}^n \, . 
\end{equation}

The extra factors $\bar{w}$ in the denominator are introduced by 
the coefficients of the $\partial_T$ derivatives and obviously 
factor out in the pure $(\partial_U)^k$ component functions. A 
direct computer calculation shows them to cancel in the other 
components, too. Without enhanced factorisation abilities this 
point is hard to show and therefore we keep the $\bar{w}$'s in the 
scheme. Incidentally, by introducing $\tau = 1/T$ the component 
functions can be more easily computed: They are simply 

\begin{equation}
\frac{1}{t^n} \partial_\tau^n \, \partial_U^m \, F \, . 
\end{equation}

We now examine the first order independence problem in detail. We 
work to lowest order, so the hat on $v$ is left out in the 
following and $R_0$ denotes the body of $R$. For charge two 
operators analyticity requires that there be functions of $(s,t)$ 
for $\D U$ and $\D w$ such that 

\begin{equation}
\Delta U \, \frac{1}{w (w R_0)} \, (\sum_{n = 0}^{4} c_{0\,n}(s,t) 
v^n) + \Delta w \, \frac{1}{w (w R_0)} \, (\sum_{n = 0}^{4} c_{1 
\,n}(s,t) v^n) = O(\frac{1}{(\hat{y}_{12} \, \hat{y}_{34})^2}) 
\label{n2First} 
\end{equation}
with $w = 1 + v$. 

Spinors occur in $\Delta U, \Delta w$ in combinations of the form 
$(\Pi \, x \, x^{-1} \, x \, \Lambda)$. Minkowski space is 
four-dimensional; a basis consists of four independent elements. 
In order to express the various $x$-triples in a given basis, we 
use 

\begin{equation}
x_i x^\dagger_j + x_j x^\dagger_i = 2 (x_i.x_j) \, \delta 
\end{equation}

to commute $x_{12}$ left of $x_{23}$ left of $x_{34}$. 
($(x_i.x_j)$ means the dot product of the associated 
four-vectors.) In this way the spinors are seen to be contracted 
on elements of the basis $\{x_{12}, x_{23}, x_{34}, x_{12} 
x^\dagger_{23} x_{34}\}$. 

There are two $\Pi_i$ and two $\Lambda_i$ so that we get a total 
of sixteen independent structures at first order in $(\Pi 
\Lambda)$. A dependence relation between $\Delta U$ and $\Delta w$ 
is a linear combination with scalar coefficients which is of a 
certain order in $\hat{y}$-singularities in all sixteen components 
separately. We restrict the analysis to the $(\Pi_1 \Lambda_2)$ 
part of the odd expansion: 

\begin{eqnarray}
\Delta U|_{\Pi_1 \Lambda_2} = & - \frac{\hat{y}_{23}}{\hat{y}_{12} 
\, \hat{y}_{34}} \, \Pi_1(x_{34}^2 \, x_{12} + x^2_{14} \, x_{23} 
+ x^2_{12} \, x_{34} + x_{12} x^\dagger_{23} x_{34})\Lambda_2 \; , 
\\ \Delta w|_{\Pi_1 \Lambda_2} = & - 
\frac{\hat{y}_{23}}{\hat{y}_{12} \, \hat{y}_{34}} \, 
\Pi_1(x_{34}^2 \, x_{12} + x^2_{12} \, x_{34} + x_{12} 
x^\dagger_{23} x_{34})\Lambda_2 
\end{eqnarray}

which we write in short as 

\begin{eqnarray}
\Delta U|_{\Pi_1 \Lambda_2} = & - \frac{\hat{y}_{23}}{\hat{y}_{12} 
\, \hat{y}_{34}} \, \Pi_1 X_U \Lambda_2 \; , \\ \Delta w|_{\Pi_1 
\Lambda_2} = & - \frac{\hat{y}_{23}}{\hat{y}_{12} \, \hat{y}_{34}} 
\, \Pi_1 X_w \Lambda_2 \; . 
\end{eqnarray}

On Minkowski space we change basis to $\{X_U, X_w, x_{12}, 
x_{34}\}$ upon which the equation (\ref{n2First}) breaks into two 
separate parts. We can omit the spinors and the $x$-vectors from 
the discussion as they have to be equal on both sides of the 
equations. This leads to the scalar equation 

\begin{equation}
- \frac{\hat{y}_{23}}{\hat{y}_{12} \, \hat{y}_{34}} \,(\sum_{i = 
0}^{4} c_n(s,t) v^n = (1+v)(s + v(1+s-t) + v^2) 
O(\frac{1}{(\hat{y}_{12} \, \hat{y}_{34})^2}) \label{scalarN2} 
\end{equation}

for both the $\Delta U$ and $\Delta W$ parts. 

Instead of the variables $\{\hat{y}_{12}, \hat{y}_{23}, 
\hat{y}_{34}\}$ we may choose the set $\{v, Y_p = 
\hat{y}_{23}/(\hat{y}_{12} \hat{y}_{34}), \hat{y}_{23}\}$. The 
Jacobian of the transform is 

\begin{equation}
J = \frac{\hat{y}_{23}^2}{(\hat{y}_{12} \, \hat{y}_{34})^3} 
(\hat{y}_{12} - \hat{y}_{34}) \label{jacobN2} 
\end{equation}

and is regular at a generic point. Given this choice, the L.H.S. 
of (\ref{scalarN2}) is independent of $\hat{y}_{23}$ and factors 
out a single power of $Y_p$. We can conclude that the same is true 
for the R.H.S. and hence the as yet unspecified term is a function 
of $s,t,v$ of maximum order $1/(\hat{y}_{12} \, \hat{y}_{34})$. It 
must be a polynomial of degree $1$ in $v$. 

It follows that the ansatz polynomial on the L.H.S. factors in the 
same way and $w (w R_0)$ cancels from the equation. The only 
solution of the regularity problem at first order is therefore the 
trivial one: 

\begin{equation}
\Delta U \, (\sum_{n = 0}^{1} g_n(s,t) v^n) + \Delta w \, (\sum_{n 
= 0}^{1} h_n(s,t) v^n) = O(\frac{1}{(\hat{y}_{12} \, 
\hat{y}_{34})^2}) \label{solN2Trivial} 
\end{equation}

This is a general solution because $\Delta U$ and $\Delta w$ are 
not more singular than $1/(\hat{y}_{12} \, \hat{y}_{34})$ in any 
of the components w.r.t. the sixteen independent structures at 
first order. By inspection, the higher order terms arising from 
the second order in $\Delta U, \Delta W$ and from the soul of 
$\hat{v}$ will also be regular. 
 
Let us state the result again: The explicit forms for the level 
one component functions in the Taylor expansion (\ref{taylExp}) 
must have the dependence on $\hat v$ indicated by the above 
equation. Hence 

\be
{D F_o\over R}=g_1 +  \hat{v} g_2 \la{du} \ee 

and 

\be
{(D+R\del_t)F_o\over \bar w R}=h_1+  \hat v h_2 \;. \la{dw} \ee 

We ought to stress that the proof makes use of certain regularity 
assumptions. Throughout the calculation we have assumed the 
$x$-scalars and in particular $s$ and $t$ to be regular. Hence 
none of the four points can be light-like separated in Minkowski 
space. This can possibly be relaxed. A necessary assumption is 
certainly that the vector basis is non-degenerate, so in 
particular the points do not coincide. For the 
$\hat{y}$-coordinates we must also demand that the points are 
distinct. For the Jacobian (\ref{jacobN2}) to be non-vanishing we 
additionally need $\hat{y}_{12} \neq \hat{y}_{34}$. 

If $\hat{y}_{34} = \alpha \hat{y}_{12}$ we can change from 
$\{\hat{y}_{12},\hat{y}_{23}\}$ to $\{\hat{y}_{23}/(\alpha \, 
\hat{y}_{12}^2), \, v\}$. In this case (\ref{scalarN2}) still has 
the same consequences: The prefactor must occur on both sides and 
drops out. The unknown part on the R.H.S. is a function of $s,t,v$ 
as before. The Jacobian of the transformation is 

\begin{equation}
J = \frac{\hat{y}_{23}}{\alpha^2 \, \hat{y}_{12}^5} ((1+\alpha) \, 
\hat{y}_{12} + 2 \, \hat{y}_{23}) 
\end{equation}

and hence is regular if the points do not coincide and 
$\hat{y}_{12}$ is not proportional to $\hat{y}_{23}$, so if there 
is more than one difference variable. The argument holds in fact 
as long as two of the $\hat{y}_{ij}$ are independent. Otherwise 
$v$ is a constant and there is nothing to discuss. 

The generalisation of the independence problem (\ref{n2First}) to 
the higher levels like $\Delta U^2,$ $\Delta U \Delta w, \, ...$ 
is obvious. We have investigated this in the same manner to lowest 
order and we find that the spinor combinations are not 
independent. Additionally, the simple argument does not apply, 
which allowed us to ignore higher orders stemming from the lower 
level terms in the Taylor expansion (\ref{taylExp}). These terms 
define a non-trivial right hand side in the equivalent of 
(\ref{n2First}) in the levels above. Due to linearity it suffices 
to find one special solution to this inhomogeneous problem which 
is to be added to the general solution of the homogeneous one. 

To analyse equation \eq{du} we begin by multiplying it by $(1+\hat 
v)R$ which gives a polynomial equation in $\hat v$. This allows us 
to express the unknown functions $g_1,g_2$ in terms of 
$a_1,a_2,a_3$, 

\be
g_1=a_{1 s} \; , \qquad g_2=a_{1 s} + a_{1 t} - a_{2 t} + a_{3 t} 
\ee 

where the literal subscripts denote partial derivatives. This 
leaves two first-order partial differential equations for 
$a_1,a_2,a_3$, 

\bea (t-s-1)a_{1s} + (t-s)a_{1t} + s(a_{2s}+a_{2t}) + 
a_2-sa_{3t}&=&0 \; , \nn\\ -a_{1s} -a_{1t}  + a_{2t} + s a_{3 s} + 
(t-1)a_{3t} + 2a_3&=&0 \; . \nn\\ \eea 

If we make the change of variables 

\bea a_1&=&\a +\c + sa_3 \; , \nn \\ a_2&=&\c +(s-t+1)a_3 \eea 

with $a_3$ unchanged, we find that $a_3$ drops out of the above 
equations altogether - it is completely undetermined. The 
equations are satisfied if 

\bea \a_s + \a_t + {\c_s}&=& 0 \, , \nn \\ t\c_t   +\c +(1-s)\a_t 
-s\a_s&=& 0 \, . \eea 

This pair of first-order coupled differential equations can be 
equivalently rewritten as a set of second-order independent ones: 

\begin{eqnarray}
s\alpha_{ss} + t\alpha_{tt} + (s+t-1)\alpha_{st} + 2(\alpha_s + 
\alpha_t) &=& 0 \; , \nonumber\\ s\gamma_{ss} + t\gamma_{tt} + 
(s+t-1)\gamma_{st} + 2(\c_s+\c_t) &=& 0\label{sameres} \;. 
\end{eqnarray}

Note that in the correlator there is still one arbitrary 
coefficient function ($a_3(s,t)$ for this choice of variables) in 
addition to the solutions to these equations. 

When carrying out this type of analysis for the higher level 
regularity problems, we find that the dependence relations between 
$\Delta U^k$ etc. introduce so many unknowns into the equations 
that no new constraints are found. We have done these calculations 
for operator weight one through three with the result that the 
only constraints arise from the first level.  


\section{$SU(2)$-covariant approach \label{CovForm}}

In this section we explain how the four-point function results 
obtained in the first part of this paper can be found in an 
independent way in a harmonic superspace formulation which 
maintains the explicit $SU(2)$ covariance. The technique is quite 
different, the main point being that here we shall keep Q 
supersymmetry (including its $SU(2)$ automorphism) manifest at 
each step. However, S supersymmetry, as well as harmonic 
analyticity will have to be checked level by level in the $\theta$ 
expansion of the four-point function. The advantage of this 
approach is the possibility to reformulate the H-analyticity 
condition in an equivalent way which will allow us to essentially 
eliminate the dependence on the G-analytic Grassmann variables. 
After this the constraints at levels 1 and 2 become sufficiently 
easy to work out. Moreover, it becomes obvious that there are no 
constraints beyond level 2.

\subsection{$SU(2)$ harmonics and Grassmann analyticity} 
 
As discussed in the previous section $N=2$ harmonic superspace is 
the product of super Minkowski space (coordinates $(x^{\m}\sim 
x^{\a\adt},\th^{\a}_i,\bar\th^{\adt i})$) and the two-sphere. 
However, in this section, rather than using explicit coordinates 
$(y,\bar y)$ for the sphere, we shall use an alternative approach 
first proposed in \cite{gikos} in which the sphere is described by 
harmonic variables $u^\pm_i$ defined as the two columns of an 
$SU(2)$ matrix; the index $i$ transforms under the (left) $SU(2)$ 
and $\pm$ under an independent (right) $U(1)$ group. The 
components of $u$ have the defining properties: 

\begin{equation}\label{harmonics}
\left(\begin{array}{cc} u^+_1 & u^-_1 \\ u^+_2 & u^-_2 \end{array} 
\right)\in SU(2)\quad \Rightarrow\quad   u^-_i =(u^{+i})^*\;, 
\quad u^{+i}u^-_i = 1  
\end{equation} 

where the $SU(2)$ indices are raised and lowered in the following 
way, $f^i = \epsilon^{ij}f_j\;,\ f_i = \epsilon_{ij}f^j$ with 
$\epsilon_{ij}$ defined by $\epsilon_{12} = -\epsilon^{12} =1$.  

The harmonic functions $f^q(u^\pm_i)$ are defined as singlets of 
the left $SU(2)$ but they are homogeneous of degree $q$ under the 
right $U(1)$, i.e. carry a charge $q$. Effectively, such functions 
live on the coset $SU(2)/U(1)\sim S^2$ and are assumed to have a 
harmonic expansion on the sphere. A powerful feature of the 
coordinateless parametrisation of $S^2$ in terms of harmonics 
$u^\pm_i$ is the possibility to write down such expansions in a 
manifestly $SU(2)$ covariant way, e.g., for $q\geq 0$: 

\begin{equation}\label{hexp}
 f^{q}(u) = \sum^\infty_{n=0} f^{(i_1\ldots i_{n+q}j_1\ldots
j_n)} u^+_{i_1}\ldots u^+_{i_{n+q}}u^-_{j_1}\ldots u^-_{j_n}\; . 
\end{equation}

The coefficients in this expansion are totally symmetric 
multispinors, i.e. irreps of $SU(2)$ of isospins $q/2 +n\;,\ 
n=0,1,\ldots$. Thus, using harmonic variables allows one to deal 
with $U(1)$ covariant objects without loosing the $SU(2)$ 
symmetry. 

With the aid of $u$ we can define Grassmann analyticity in an 
$SU(2)$-covariant way. We split the Grassmann variables 
$\theta^{\alpha}_i, \bar\theta^{\dot\alpha i}$ into two $U(1)$ 
projections, 
\begin{equation}
  \theta^{\pm\alpha} = u^\pm_i\theta^{i\alpha},\quad 
\bar\theta^{\pm\dot\alpha} = u^\pm_i\bar\theta^{i\dot\alpha} \;, 
\end{equation}
still maintaining the $SU(2)$ invariance. We then define a 
G-analytic function $\f$ on harmonic superspace to be one which 
satisfies 

\be
D_{\a}^{+}\f=\bar D_{\adt}^{+}\f=0 \ee 

where the $U(1)$ projections of the supercovariant derivatives are 
defined in a similar way. These constraints are solved by 

\be
\f=\f(x_A,\th^{+\a},\bar\th^{=\adt}) \ee 

where 

\begin{equation}
  x_A^{\alpha\dot\alpha}\ = \  x^{\alpha\dot\alpha} -4i \theta^{(i\, \alpha}
  \bar\theta^{j)\, \dot\alpha}u^+_iu^-_j 
\end{equation}

with $x^{\alpha\dot\alpha} = x^\mu \sigma_\mu^{\alpha\dot\alpha}$. 
Clearly G-analyticity is a Q-supersymmetric notion because it 
involves the supercovariant derivatives. 

As explained in the previous section, complexified analytic 
superspace can be defined as a coset space of the complex 
superconformal group. Although this is not possible in real 
spacetime we can nevertheless define a representation of the Lie 
superalgebra of $SU(2,2|2)$ on $G$-analytic fields. For all 
practical purposes this amounts to taking infinitesimal $SU(2)$ 
transformations (parameters $\lambda^{jk}$) to have the form 
$u^\pm_i$: 

\begin{equation}\label{su2c}
  \delta u^+_i = (\lambda^{jk}u^+_ju^+_k)\;u^-_i\;, \quad \delta u^-_i = 
0\;. 
\end{equation}

More generally, it is sufficient to treat the transformations of 
the harmonic variables $u^\pm_i$ with generators $K,D,R,S,I$ as 
active ones. For instance, a harmonic function $f^{(p)}(u)$ of  
weight $p$ will transform as follows: 

\begin{eqnarray}
  \delta f^{(p)}(u) &=&  f'{}^{(p)}(u)  - f^{(p)}(u)\nonumber\\
   &=& -(\lambda^{ij}u^+_i u^+_j)u^-_k{\partial\over\partial u^+_k}f^{(p)}(u) +
   p(\lambda^{ij}u^-_i u^+_j)f^{(p)}(u)\;,
\end{eqnarray}

so that the non-unitary transformation appears in the form of a 
derivative of a function of unitary harmonics. 

Henceforth we shall be dealing exclusively with G-analytic fields 
so that we can replace $x_A$ by $x$ without loss of clarity. The 
actions of the supersymmetry transformations on the coordinates 
are given by 

\bea \d_Q x^{\a\adt}&=&-4iu_i^-(\e^{i\a}\bar\th^{+\adt} + 
\th^{+\a}\bar\e^{i\adt}) \nn\\ 
\d_Q\th^{+\a,\adt}&=&u_i^{+}\e^{i\a,\adt}\nn \\ \d_Q u_i^{\pm}&=&0 
\la{Qsusy} \eea 

and \cite{fradkin} 

\begin{eqnarray}
\delta_S x^{\alpha\dot\alpha} &=& 4i 
(x^{\alpha\dot\beta}\bar\theta^{+\dot\alpha}\bar\eta^i_{\dot\beta} 
-x^{\dot\alpha\beta}\theta^{+\alpha}\eta^i_\beta)u^-_i\nonumber\\ 
\delta_S\theta^{+\alpha} &=& -2i (\theta^+)^2\eta^{\alpha i}u^-_i 
+ x^{\alpha\dot\beta}\bar\eta^i_{\dot\beta}u^+_i\nonumber\\  
\delta_S\theta^{-\alpha} &=&  
4i\eta^i_\beta\theta^{-\beta}(\theta^{-\alpha}u^+_i-\theta^{+\alpha}u^-_i) 
+\bar\eta^i_{\dot\beta}(x^{\alpha\dot\beta} 
+4i\theta^{-\alpha}\bar\theta^{+\dot\beta})u^-_i\nonumber\\ 
\delta_S u^+_i &=& [4i (\theta^{+\alpha}\eta^i_\alpha 
+\bar\eta^i_{\dot\alpha}\bar\theta^{+\dot\alpha})u^+_i]u^-_i\nonumber\\ 
\delta_S u^-_i &=& 0 \label{7.9.1c} 
\end{eqnarray} 

($\delta_S\bar\theta^\pm$ are obtained by conjugation). From this 
one can compute the action of the rest of the superconformal 
algebra by commuting Q and S supersymmetry transformations.

\subsection{The hypermultiplet}

In this subsection we give the harmonic formulation of the 
hypermultiplet describing an $SU(2)$ doublet of scalars $f^i(x)$ 
and a pair of Weyl (complex) spinors $\psi_\alpha(x),\ 
\bar\xi^{\dot\alpha}(x)$ on shell. Off shell it can only exist 
with an infinite set of auxiliary fields \cite{nogo}. Such a set 
is naturally provided by the G-analytic superfield of $U(1)$ 
charge $+1$: 

\begin{equation}\label{q+}
  q^+(x,\theta^+,\bar\theta^+,u) = F^+(x,u) + \theta^{+\alpha}\Psi_\alpha(x,u) +
\bar\theta^+_{\dot\alpha} \bar\Xi^{\dot\alpha}(x,u) + ... 
+(\theta^+)^2(\bar\theta^+)^2 P^{-3}(x,u)\;. 
\end{equation}

The components of this $\theta^+$ expansion are harmonic functions 
with infinite expansions on $S^2$ (see (\ref{hexp})): 

\begin{eqnarray}
 F^+(x,u) &=& f^i(x)u^+_i + f^{(ijk)}(x)u^+_iu^+_ju^-_k + \ldots \nonumber\\
 \Psi_\alpha(x,u) &=& \psi_\alpha(x) + \psi^{(ij)}_\alpha(x))u^+_iu^-_j
  + \ldots \nonumber\\ 
 &&  \ldots \nonumber\\
 P^{-3}(x,u) &=& p^{(ijk)}(x)u^-_iu^-_ju^-_k + \ldots 
\end{eqnarray}

The coefficients in these expansions are ordinary fields belonging 
to different $SU(2)$ representations. All of them, with the 
exception of the physical fields $f^i(x), \psi_\alpha(x), 
\bar\xi^{\dot\alpha}(x)$ are auxiliary, so they should vanish on 
shell. We need a way to write down a supersymmetric on-shell 
constraint on the G-analytic superfield 
$q^+(x,\theta^+,\bar\theta^+,u)$. 

The key to this on-shell constraint is provided by the notion of 
harmonic (H-)analyticity. The harmonic coset $SU(2)/U(1)$ has two 
real (or one complex) dimensions, to which correspond the 
following harmonic derivatives: 

\begin{eqnarray}
 \partial^{++} = u^+_i{\partial\over\partial u^-_i}\ &\Rightarrow&\ 
\partial^{++}u^+_i = 0\;, \ \ \partial^{++}u^-_i = u^+_i\;, \nonumber\\
 \partial^{--} = u^-_i{\partial\over\partial u^+_i} \ &\Rightarrow&\ 
\partial^{--}u^+_i = u^-_i\;, \ \ \partial^{--}u^-_i = 0\;. \label{harder}
\end{eqnarray}

These are Cartan's covariant derivatives on the coset. In our 
context this simply means that they preserve the defining 
condition $u^{+i}u^-_i = 1$. To them one may add the 
charge-counting operator 

\begin{equation}\label{D0}
  \partial^0 = u^+_i{\partial\over\partial u^+_i} 
- u^-_i{\partial\over\partial u^-_i}\ \Rightarrow \  
\partial^0u^\pm_i = \pm u^\pm_i\;. 
\end{equation}

By definition all harmonic functions are eigenfunctions of 
$\partial^0$, $\partial^0 f^q(u) = q f^q(u)$.  

An important point is that the three covariant derivatives above 
form an $SU(2)$ algebra: 

\begin{equation}
\left[\partial^{++},\partial^{--}\right] = \partial^0\;,\ \ \ 
\left[\partial^0,\partial^{\pm\pm}\right] = \pm 
2\partial^{\pm\pm}\;. \label{3.71} 
\end{equation} 

They can be regarded as the generators of right $SU(2)_R$ 
rotations acting on the indices $\pm$ of the harmonics $u^\pm_i$. 
Thus, $\partial^{++}$ is the raising and $\partial^{--}$ the 
lowering operator of $SU(2)_R$ (see (\ref{harder})). This 
observation suggests the way to define short harmonic functions as 
highest weights of irreps of $SU(2)_R$. Thus, depending on the 
value of the $U(1)$ charge, the condition 

\begin{equation}\label{ana}
  \partial^{++}f^q(u)=0 \quad \Rightarrow \quad
  \left\{\begin{array}{ll} f^q(u)=0, \ q<0 \\
  f^q(u)= u^+_{i_1}\ldots u^+_{i_q} f^{(i_1\ldots i_q)},
  \ q \geq 0\end{array} \right.
\end{equation}

has either a trivial solution or defines an irrep of isospin 
$q/2$. This property is a direct consequence of the general form 
(\ref{hexp}) of the harmonic expansion on $S^2$ and of the action 
of $\partial^{++}$ on the harmonics (\ref{harder}).   

An alternative interpretation of the condition (\ref{ana}) is that 
of harmonic (H-)analyticity. Let us introduce stereographic 
coordinates on the sphere (see \cite{book} for more detail): 

\begin{equation}
\left(\begin{array}{cc} u^+_1 & u^-_1 \\ u^+_2 & u^-_2 \end{array} 
\right) = {1 \over \sqrt{1+y\bar y}} \left(\begin{array}{cc} 1 & 
-\bar y \\ y & 1\end{array} \right) \;. \label{4.1.3} 
\end{equation} 

Our harmonic functions $f^q(u)$ are by definition eigenfunctions 
of the charge ``operator" $\partial^0$: 

\begin{equation}
\partial^0 f^{q}(y,\bar y) = qf^{q}(y,\bar y)\;. \label{4.2.1} 
\end{equation}

In this parametrisation the covariant derivative $\partial^{++}$ 
becomes     

\begin{equation}
\partial^{++} = -(1+y\bar y){\partial\over \partial \bar 
y} - {y\over 2}\partial^0 \;. \label{4.5.1} 
\end{equation} 

In these terms eq. (\ref{ana}) takes the form of a (covariant) 
harmonic analyticity condition: 

\begin{equation}
  {\partial f^{q}\over
\partial\bar y} + {qy\over 2(1+y\bar y)}f^{q} = 0\;. 
\end{equation}

It admits the general solution $f^{q}(y,\bar y) = (1+y\bar 
y)^{-{q\over 2}} f_0(y)$ where $f_0(y)$ is an arbitrary 
holomorphic function. Remembering that we are looking for 
solutions globally defined on the sphere it is not hard to show  
that for $q<0$ the only solution is $f_0=0$ and for $q\geq 0$ 
$f_0(y)$ must be a polynomial of degree $q$ whose $SU(2)$ 
covariant form is given in (\ref{ana}). 

It should be stressed that the above H-analytic harmonic functions 
are regular, i.e. well-defined on the whole of $S^2$. In practice 
one has also to deal with singular harmonic functions. A typical 
example we shall encounter in what follows is the harmonic 
distribution 

\begin{equation}
{1\over (12)} \label{4.8.7} 
\end{equation} 

where 

\begin{equation}
(12) \equiv u^{+i}_1u^+_{2i} = {y_2-y_1\over \sqrt{(1+y_1\bar 
y_1)(1+y_2\bar y_2)}}\;. \label{4.8.8} 
\end{equation} 

At first sight, it is a function of $u^+$ only, therefore one 
would expect $\partial^{++}_1(12)^{-1} = 0$. However, this 
distribution is singular at the point $u_1=u_2$, so it should be 
differentiated with care: 

\begin{eqnarray}
\partial^{++}_1 {1\over (12)} &=& (1+y_1\bar 
y_1)^{3/2} (1+y_2\bar y_2)^{1/2} {\partial\over \partial\bar 
y_1}{1\over y_1-y_2} \nonumber\\ &=& (1+y_1\bar y_1)^2 i\pi 
\delta(y_1-y_2) \nonumber\\ &=& (12^-)\delta(u_1,u_2)\;, 
\label{4.9.1} 
\end{eqnarray} 

where we have used the well-known relation 

\begin{equation}
{\partial\over \partial\bar y}{1\over y} = i\pi \delta(y)\;. 
\label{4.9.2} 
\end{equation} 

Note that the factor $(12^-)\equiv u^{+i}_1u^-_{2i}$ is needed for 
keeping the balance of charges on both sides of eq. (\ref{4.9.1}) 
in the $S(2)$ covariant notation. 

The conclusion from the above discussion is that the condition 
(\ref{ana}) is, on the one hand, the definition of a highest 
weight of an $SU(2)$ irrep and, on the other hand, a harmonic 
analyticity condition on the sphere. This type of condition can 
easily be supersymmetrised in order to be applied to superfields 
such as the hypermultiplet $q^+$ (\ref{q+}) and we obtain the 
following operator invariant under Q supersymmetry: 

\begin{equation}
D^{++} = \partial^{++} -2i\theta^{+\alpha}\bar\theta^{+\dot\alpha} 
\partial_{\alpha\dot\alpha}
\end{equation} 

where $\partial_{\alpha\dot\alpha} \equiv 
\sigma^\mu_{\alpha\dot\alpha}\partial/\partial x^\mu$. 
 
Now, let us impose the (supercovariant) H-analyticity condition 

\begin{equation}\label{Hanq}
  D^{++} q^+(x,\theta^+,\bar\theta^+,u) = 0\;.
\end{equation}

Inserting the expansion (\ref{q+}) into eq. (\ref{Hanq}), we 
obtain a set of harmonic differential equations which are solved 
just like eq. (\ref{ana}). The result is the short (on-shell) 
hypermultiplet 

\begin{equation}\label{5.2.12}
  q^+ = f^i(x)u^+_i +\theta^{+\alpha}\psi_\alpha(x) 
+\bar\theta^+_{\dot\alpha}\bar\xi^{\dot\alpha}(x) 
+2i\theta^{+\alpha}\bar\theta^{+\dot\alpha} 
\partial_{\alpha\dot\alpha}f^i(x)u^-_i  
\end{equation}

where all the auxiliary fields have been eliminated and the 
remaining physical ones put on shell, 
$$
\square f^i(x)=\dslash\psi =\dslash\bar\xi=0\;. 
$$

So, in the case of the hypermultiplet the combination of G- and 
H-analyticities results in an on-shell superfield. Note that this 
result crucially depends on the $U(1)$ charge of the G-analytic 
superfield. For example, a superfield 
$L^{++}(x,\theta^+,\bar\theta^+,u)$ of charge $+2$ subject to the 
same H-analyticity condition 

\begin{equation}\label{linm}
  D^{++}L^{++}=0
\end{equation}

describes an off-shell multiplet (the linear or tensor multiplet 
consisting of a triplet of real scalars, a divergenceless real 
vector, a Majorana spinor and a complex auxiliary field). For 
charges $\geq+3$ the H-analyticity condition simply cuts off the 
tail of auxiliary fields without imposing any constraints on the 
remaining physical fields. On the contrary, for charges $\leq 0$ 
the condition is too strong and only admits a trivial solution.  

A very important observation is that the H-analyticity conditions 
(\ref{Hanq}) or (\ref{linm}) admit an equivalent form in terms of 
the harmonic derivative $\partial^{--}$. Remembering that 
$\partial^{++}$ and $\partial^{--}$ are the raising and lowering 
operators of $SU(2)_R$ and that the H-analyticity condition 
$\partial^{++}f^q(u)=0$ defines the highest weight of an  
$SU(2)_R$ irrep of isospin $q/2$ (dimension $q+1$), we immediately 
see the equivalence relation 

\begin{equation}\label{equiv}
  \partial^{++}f^q(u)=0\quad \Leftrightarrow \quad 
(\partial^{--})^{q+1}f^q(u)=0 
\end{equation}

(alternatively, it can be derived by inspecting the harmonic 
expansion (\ref{hexp})). The supersymmetric version of the new 
form of the H-analyticity condition involves the operator 

\begin{equation}\label{d--}
D^{--} = \partial^{--} - 
2i\theta^{-\alpha}\bar\theta^{-\dot\alpha} 
\partial_{\alpha\dot\alpha} + 
\theta^{-\alpha}{\partial\over\partial\theta^{+\alpha}} + 
\bar\theta^{-\dot\alpha}{\partial\over\partial\bar\theta^{+\dot\alpha}}\; 
. \label{3.70} 
\end{equation}

There is a crucial difference between these two conditions, which 
we shall heavily exploit in what follows. The point is that 
singular harmonic functions of the type (\ref{4.8.7}) give rise to 
delta-type singularities under $\partial^{++}$ (see  
(\ref{4.9.1})), whereas they can be differentiated as ordinary 
functions by $\partial^{--}$, e.g. 

\begin{equation}\label{ordder}
  \partial^{--}{1\over (12)} = -{(1^-2)\over (12)^2}\;.
\end{equation}

The explanation is that in the former case we deal with a 
derivative of the type $\partial/\partial\bar y\; y^{-1} = i\pi 
\delta(y)$ and in the latter $\partial/\partial y\; y^{-1} = 
-y^{-2}$. 
 
In the rest of this section we shall examine the non-trivial 
implications of H-analyticity combined with the requirement of 
superconformal covariance for correlation functions of charge 
$+2$. For that purpose we shall need the superconformal 
transformation properties of the superfields and operators we have 
introduced. The transformation law of the harmonic derivatives 
$D^{++}$ and $D^{--}$ can be found using Cartan's coset scheme 
\cite{book} (or checked directly \cite{fradkin}): 

\begin{eqnarray}
\delta D^{++} &=&-\Lambda^{++}D^0\;,\nonumber\\ \delta D^{--} &=& 
-(D^{--}\Lambda^{++})D^{--} \label{7.9.2} 
\end{eqnarray}  

where 

\begin{equation}
D^{++}\Lambda = \Lambda^{++}\;, \ \ \ D^{++}\Lambda^{++} =0 
\label{7.10.2} 
\end{equation} 

and 

\begin{equation}\label{7.13.22}
 \Lambda = a + k_{\alpha\dot\alpha}x^{\alpha\dot\alpha} 
+\lambda^{ij}u^+_iu^-_j + 4i (\theta^{+\alpha}\eta^i_\alpha 
+\bar\eta^i_{\dot\alpha}\bar\theta^{+\dot\alpha})u^-_i\end{equation} 

is the superconformal weight factor. For completeness, besides the 
S supersymmetry parameter $\eta$ we have also included those of 
dilation $a$, conformal boosts $k^\mu$ and $SU(2)_C$ 
$\lambda^{ij}$. Then it is not hard to check that the  
H-analyticity condition  

\begin{equation}\label{equivq}
  D^{++}q^+ =0\quad \Leftrightarrow \quad 
(D^{--})^2q^+=0 
\end{equation}

is covariant if the hypermultiplet transforms with superconformal 
weight $+1$: 

\begin{equation}
\delta q^+ = -\lambda\cdot\partial q^+ + \Lambda q^+ 
\label{7.13.2} 
\end{equation} 

(here $-\lambda\cdot\partial$ denotes the coordinate 
transformations). Similarly, the linear multiplet subject to the  
H-analyticity condition 

\begin{equation}\label{equivL}
  D^{++}L^{++} =0\quad \Leftrightarrow \quad 
(D^{--})^3L^{++}=0 
\end{equation}

should have weight $+2$: 

\begin{equation}
\delta L^{++} = -\lambda\cdot\partial L^{++} + 2\Lambda L^{++}\;. 
\label{Ltransf} 
\end{equation}  

\subsection{Two- and three-point functions}

The simplest example of a two-point function is the hypermultiplet 
propagator 

\begin{equation}
  G^{(1,1)} (1\vert 2) = \langle \tilde q^+(1)q^+(2)\rangle
\end{equation}

where the superscript $(1,1)$ indicates the $U(1)$ charges at the 
two points and $\tilde{}$ is a special conjugation on $S^2$ 
preserving G-analyticity \cite{gikos}. It is defined as the 
Green's function of the field equation (\ref{Hanq}):   

\begin{equation}\label{14.1.3}
D^{++}_1 G^{(1,1)} (x_1,\theta^+_1,u_1\vert x_2,\theta^+_2,u_2) = 
\delta^4(x_1-x_2)(\theta^+_1-(u^+_1u^-_2)\theta^+_2)^4(u^-_1u^+_2)\delta 
(u_1,u_2)\;. 
\end{equation}

Naturally, like the hypermultiplet superfield $q^+$ itself, the 
Green's function should be G-analytic. The right-hand side of eq. 
(\ref{14.1.3}) is the complete delta-function of the G-analytic 
superspace (as in eq. (\ref{4.9.1}), the factors $(u^+_1u^-_2)$ 
and $(u^+_1u^-_2)$ maintain the balance of $U(1)$ charges). 
Throughout this paper we assume that all the correlation functions 
are considered away from the coincident points where they usually 
have singularities. In this case the right-hand side of eq. 
(\ref{14.1.3}) just vanishes: 

\begin{equation}\label{hanG}
  D^{++}_1 G^{(1,1)} (1\vert 2) = 0 \ \ \mbox{for points $1\neq 2$.}
\end{equation}

The same is of course true if we replace $D^{++}_1$ by $D^{++}_2$. 
In other words, this two-point function is H-analytic away from 
the singular point.  

Another basic property of the hypermultiplet propagator is 
superconformal covariance. According to the transformation law 
(\ref{7.13.2}) of the hypermultiplet itself, the propagator 
transforms as follows: 

\begin{equation}
\delta G^{(1,1)}(1\vert 2) = -\lambda\cdot\partial 
G^{(1,1)}(1\vert 2)  + (\Lambda(1) + \Lambda(2)) G^{(1,1)}(1\vert 
2) \;. \label{7.13.20} 
\end{equation}

The combination of H-analyticity and the conformal properties of 
the propagator allow us to find the explicit expression for 
$G^{(1,1)}$. We start by examining the leading component of this 
superfield 

\begin{equation}\label{lcom}
  g^{(1,1)}(x^2_{12},u_1,u_2) = 
G^{(1,1)}(\theta^+_1=\theta^+_2=0)\;. 
\end{equation}

Here we have taken into account translation and Lorentz invariance 
which tell us that the function must depend on the space-time 
invariant $x^2_{12}\equiv (x_1-x_2)^2$. In the absence of 
$\theta^+$ the harmonic derivative $D^{++} \equiv 
\partial^{++}$, so the H-analyticity condition (\ref{hanG}) simply tells 
us that $g^{(1,1)}$ must be linear in $u^+_1$ (recall 
(\ref{ana})). At the same time it is an $SU(2)$ invariant, so the 
index $i$ of $u^+_{1i}$ must be contracted with the other harmonic 
variable $u^\pm_2$. Given the charges $+1$ at both points, we 
conclude that the only such invariant combination of harmonics is 
$(12)\equiv u^{+i}_i u^+_{2i}\;$. So, $g^{(1,1)}$ is reduced to 

\begin{equation}\label{inhere}
  g^{(1,1)} = (12)g(x^2_{12})\;.
\end{equation}

The remaining function $g(x^2_{12})$ can be most easily determined 
by making use of the dilation part of the conformal group. The 
first component of the superfield $q^+$ is the physical scalar 
$f^i(x)$ which has conformal weight 1, and so does the leading 
term in the hypermultiplet propagator. The harmonic factor in 
(\ref{inhere}) is weightless, so we conclude that $g = 
C/x^2_{12}$. The constant $C$ can be fixed by comparing with the 
standard scalar propagator and the result is 

\begin{equation}\label{leco}
  g^{(1,1)} ={1\over 
4i\pi^2} {(12)\over x^2_{12}} \;. 
\end{equation}
 
Now, the less trivial part of the determination of the propagator 
$G^{(1,1)}$ is completing it to a full superfield, i.e. restoring 
the dependence on $\theta^+_{1,2}$. Here we shall use a trick 
which will prove very useful in the study of the four-point 
correlator in the next subsection. The two-point function is 
supposed invariant under Q supersymmetry (\ref{Qsusy}), which acts 
as a shift of the Grassmann variables: 

\begin{equation}\label{epsii} 
  ({\theta^{+\alpha,\dot\alpha}_{1,2}})'=\theta^{+\alpha,\dot\alpha}_{1,2} 
+  u^+_{1,2i}\epsilon^{i\alpha,\dot\alpha}\;. 
\end{equation}

It is then clear that by making a finite Q supersymmetry 
transformation with parameter 

\begin{equation}\label{epsi}
  \epsilon^{i\alpha,\dot\alpha} = {u^{+i}_2\over(12)}\theta^{+\alpha,\dot\alpha}_1
 - {u^{+i}_1\over(12)}\theta^{+\alpha,\dot\alpha}_2
\end{equation} 

we can eliminate both $\theta^+_1$ and $\theta^+_2$: 

\begin{equation}\label{Qfr}
\mbox{Q frame:}\qquad  
({\theta^{+\alpha,\dot\alpha}_{1}})'=({\theta^{+\alpha,\dot\alpha}_{2}})'= 
0\;. 
\end{equation}

In this frame the two-point function becomes independent of the 
Grassmann variables. In other words, it coincides with its leading 
component (\ref{leco}), $G^{(1,1)}\vert_Q \equiv g^{(1,1)}$. Then 
we can go back to the original frame by performing the same finite 
Q supersymmetry transformations on the remaining 
coordinates.\footnote{As a simpler example of this trick, consider 
translation invariance for a set of two space-time points 
$x_1,x_2$. By means of the finite translation $P:\ x_2'= x_2 + a = 
0$ we can go to a $P$ frame in which only $x_1$ survives. Then, to 
restore manifest invariance, we make the same shift on $x_1$: 
$x'_1 = x_1 + a = x_1-x_2\equiv x_{12}$.} Such a transformation 
only affects the difference $x_{12}$ and gives 

\begin{equation}\label{hataa}
  \hat x^{\alpha\dot\alpha}_{12} = x^{\alpha\dot\alpha}_{12} + {4i\over (12)}[(1^-2) \theta^+_1 \bar\theta^+_1 + 
(2^-1) \theta^+_2 \bar\theta^+_2 +  \theta^+_1 \bar\theta^+_2 +  
\theta^+_2 \bar\theta^+_1]^{\alpha\dot\alpha}\;. 
\end{equation}

By construction, this modified coordinate difference is invariant 
under Q supersymmetry, which can be easily verified using 
(\ref{Qsusy}). So, to find out the $\theta^+_{1,2}$ dependence of 
the hypermultiplet propagator, it is enough to replace $x_{12}$ by 
$\hat x_{12}$: 

\begin{equation}\label{propa}
  G^{(1,1)}(1\vert 2) ={1\over 
4i\pi^2} {(12)\over\hat x^2_{12}}\;. 
\end{equation}

In deriving this two-point function we have only used the dilation 
part of the superconformal group. In fact, since the result is 
unique, it is guaranteed to have the right superconformal 
properties (\ref{7.13.20}) of the propagator (this can also be 
checked directly). Further, so far we have only solved the 
H-analyticity constraint (\ref{hanG}) at the lowest (leading) 
order of the $\theta^+$ expansion. One might try to argue that 
since the left-hand side of eq. (\ref{hanG}) is itself an 
invariant of Q supersymmetry, it is sufficient to check 
H-analyticity in the Q frame (i.e., in the absence of $\theta^+$). 
However, this argument is not safe here. Indeed, in the expansion 
of the two-point function there are harmonic singularities of pole 
type (e.g., $(12)^{-1}$), on which the operator $\partial^{++}$ 
creates a delta-type singularity (recall (\ref{4.9.1})). In such a 
situation we will not be allowed to use the supersymmetry 
parameter (\ref{epsi}) which itself contains harmonic poles. A 
safe way to extend H-analyticity to all orders in the $\theta^+$ 
expansion by means of the transformation (\ref{epsi}) is to use 
the alternative form of the H-analyticity constraint involving 
$D^{--}$ (see (\ref{equivq})). We shall come back to this 
important point in the next subsection.  

Knowing the hypermultiplet propagator, we can easily predict the 
general form of correlators of two or three composite operators 
made out of hypermultiplets. Take, for instance, the two-point 
function 

\begin{equation}\label{2ptco}
 G^{(2,2)}(1\vert 2) = \langle \mbox{Tr}\left(\tilde q^+(1)\right)^2
 \mbox{Tr}\left(q^+(2)\right)^2\rangle\;.
\end{equation}

Note that it has charges $+2$ at each point matching the number of 
elementary hypermultiplets in each composite operator.  One can 
imagine this correlator in the context of $N=4$ super-Yang-Mills 
theory, where a hypermultiplet in the adjoint representation of 
the gauge group interacts with the $N=2$ super-Yang-Mills gauge 
potential. Since the $N=4$ theory is finite (conformally 
invariant), we can demand that the correlator (\ref{2ptco}) be 
superconformally covariant, 

\begin{equation}\label{trG2}
 \delta G^{(2,2)}(1\vert 2) = -\lambda\cdot\partial 
G^{(2,2)}(1\vert 2)  + 2(\Lambda(1) + \Lambda(2)) G^{(2,2)}(1\vert 
2) \;. 
\end{equation}

In addition to this, the correlator should be H-analytic. Indeed, 
let us differentiate it with the harmonic derivative 
$D^{++}$.\footnote{The traces in (\ref{2ptco}) make the composite 
operators gauge invariant, so we can use a flat $D^{++}$ (no gauge 
connection).} Since $D^{++}$ is the operator of the free field 
equation (\ref{q+}) for the hypermultiplet, one can argue that 
such a differentiation will give rise to a Schwinger-Dyson 
equation for the correlator: 

\begin{equation}\label{SD}
  D^{++}_1G^{(2,2)}(1\vert 2) = \mbox{contact terms}\;.
\end{equation}

Since the composite operators are bilinear in this case (charges 
$+2$), equation (\ref{SD}) can also be interpreted as a Ward 
identity. Indeed, the bilinears $\tilde q^+\tilde q^+$, $\tilde 
q^+ q^+$ and $q^+q^+$ are the currents of an extra $SU(2)$ 
symmetry of the $N=4$ theory realised in terms of $N=2$ 
superfields\footnote{In fact, this symmetry is the visible part of 
the full $SU(4)$ R symmetry of the $N=4$ theory.}. So, in this 
case eq. (\ref{SD}) corresponds to the current conservation law. 
In the context of this paper we treat contact terms as zeros, so 
eq. (\ref{SD}) takes the form of an H-analyticity condition: 

\begin{equation}\label{hanG2}
  D^{++}_1 G^{(2,2)} (1\vert 2) = 0 \ \ \mbox{for points $1\neq 2$}
\end{equation}

(and similarly at point 2). Since the product of two H-analytic 
functions is H-analytic as well, we immediately find an obvious 
solution to this constraint as the square of the hypermultiplet 
propagator, 

\begin{equation}\label{crrr}
  G^{(2,2)} (1\vert 2) = C {(12)^2\over\hat x^4_{12}}
\end{equation}

where $C$ is a constant. In fact, this is the general solution. 
The argument is as in the case of the propagator. One first 
examines the leading component 
$G^{(2,2)}(\theta^+_1=\theta^+_2=0)$. The constraint (\ref{hanG2}) 
fixes the harmonic dependence since the combination $(12)^2$ is 
the only $SU(2)$ invariant of charges $(2,2)$ annihilated by 
$\partial^{++}$. The dependence on $x^2_{12}$ is determined by 
simple dilation covariance. Finally, with two G-analytic Grassmann 
variables $\theta^+_{1,2}$ we already know that the complete 
$\theta^+$ dependence is fixed by Q supersymmetry alone, by just 
putting a hat on $x^2_{12}$. It should be mentioned that the above 
considerations cannot predict the value of the constant in 
(\ref{crrr}). In principle, it might receive quantum corrections 
at each level of perturbation theory, but it can be shown that 
this type of correlator is protected by a non-renormalisation 
theorem  \cite{hw1,hsw}.  

Next we turn to three-point correlators. As an example, take the 
correlator of three currents, i.e. bilinears made out of 
hypermultiplets: 

\begin{equation}\label{3ptco}
 G^{(2,2,2)}(1\vert 2\vert 3) = 
\langle \mbox{Tr}\left(\tilde q^+(1)\right)^2 
\mbox{Tr}\left(\tilde q^+(2)q^+(2)\right) 
\mbox{Tr}\left(q^+(3)\right)^2\rangle 
\end{equation}
and subject to the requirements of H-analyticity 
\begin{equation}\label{hanG3}
  D^{++}_1  G^{(2,2,2)}(1\vert 2\vert 3) = 0 \ \ 
\mbox{for points $1\neq 2\neq 3$} 
\end{equation}
and of superconformal covariance 
\begin{equation}\label{trG3}
 \delta G^{(2,2,2)}(1\vert 2\vert 3) = -\lambda\cdot\partial 
G^{(2,2,2)}(1\vert 2\vert 3)  + 2(\Lambda(1) + \Lambda(2) + 
\Lambda(3)) G^{(2,2,2)}(1\vert 2\vert 3) \;. 
\end{equation}

Just as for $G^{(2,2)} (1\vert 2)$ above, it is obvious that the 
product of three propagators 

\begin{equation}\label{crr3}
  G^{(2,2,2)}(1\vert 2\vert 3) = C {(12)\over\hat x^2_{12}}
{(23)\over\hat x^2_{23}}{(31)\over\hat x^2_{31}} 
\end{equation}

satisfies both requirements. To prove its uniqueness, we argue as 
follows. Firstly, at the lowest level in the $\theta^+$ expansion 
there is a single $SU(2)$ invariant combination of the three 
harmonics with the right charges and vanishing under $D^{++}_1$, 
namely $(12)(23)(31)$. Secondly, the space-time dependence is now 
determined by the full conformal group (and not just dilations, as 
for two points). It is well-known that there exists no conformal 
invariant made out of three space-time variables, therefore the 
product $x^{-2}_{12}x^{-2}_{23}x^{-2}_{31}$  is the only function 
with the required conformal properties. Finally, we have to show 
that putting hats on the $x$'s gives the unique completion of the 
leading component to a full superfield. Before we did this by 
using the Q frame (\ref{Qfr}) in which the two Grassmann variables 
had been eliminated. Now we have three $\theta^+$'s, and the Q 
supersymmetry parameter $\epsilon^i$ alone is not enough to shift 
away all of them. This time we have to invoke S supersymmetry as 
well. Looking at the transformation law of $\theta^+$ in 
(\ref{7.9.1c}) we see that S supersymmetry acts essentially as a 
shift (although non-linear), provided that the matrix 
$x^{\alpha\dot\alpha}$ is invertible. Then the combination of Q 
and S supersymmetry makes it possible to find a   

\begin{equation}\label{QSfr}
\mbox{Q{\&}S frame:}\qquad  
({\theta^{+\alpha,\dot\alpha}_{1}})'=({\theta^{+\alpha,\dot\alpha}_{2}})'= 
({\theta^{+\alpha,\dot\alpha}_{3}})'=0 
\end{equation}

in which there are no $\theta^+$'s left.\footnote{In fact, the S 
supersymmetry parameter $\eta^i$ is an $SU(2)$ doublet, just as 
$\epsilon^i$. Using both of them we can shift away up to four 
$\theta^+$, as we shall do in the four-point case.} So, if there 
existed another superfield completion of the leading component 
above, their difference would be a nilpotent (i.e., proportional 
to $\theta^+$) superconformal covariant. But such an object would 
vanish in the Q{\&}S frame, therefore it must vanish in any frame. 
So, H-analyticity and superconformal covariance can predict the 
form of the three-point correlator up to a constant factor. Once 
again, it turns out protected by a non-renormalisation theorem 
\cite{hw1,hsw}.

\subsection{Four-point correlators}

\subsubsection{Preliminaries}

The main subject of interest in this paper are four-point 
correlators of hypermultiplet bilinears of the type, e.g., 

\begin{equation}\label{4ptco}
 G^{(2,2,2,2)}(1\vert 2\vert 3\vert 4) = 
\langle \mbox{Tr}\left(\tilde 
q^+(1)\right)^2\mbox{Tr}\left(q^+(2)\right)^2 
\mbox{Tr}\left(\tilde 
q^+(3)\right)^2\mbox{Tr}\left(q^+(4)\right)^2\rangle 
\end{equation}
satisfying the requirements of H-analyticity 
\begin{equation}\label{hanG4}
  D^{++}_1 G^{(2,2,2,2)}(1\vert 2\vert 3\vert 4) = 0 \ \ 
\mbox{for points $1\neq 2\neq 3\neq 4$} 
\end{equation}

and of superconformal covariance 

\begin{equation}\label{trG4}
 \delta G^{(2,2,2,2)}(1\vert 2\vert 3\vert 4) = -\lambda\cdot\partial 
G^{(2,2,2,2)}(1\vert 2\vert 3\vert 4)  + 2(\Lambda(1) + \Lambda(2) 
+ \Lambda(3)+ \Lambda(4)) G^{(2,2,2,2)}(1\vert 2\vert 3\vert 4) 
\;. 
\end{equation}

Compared to the two- and three-point cases above, the structure of 
the four-point correlator is considerably richer, for two main 
reasons which can be seen at the lowest level in the $\theta^+$ 
expansion. Firstly, now there exist three independent harmonic 
combinations satisfying (\ref{hanG4}): 

\begin{equation}\label{3str}
  (12)^2(34)^2\;,\ (14)^2(23)^2\;,\ (12)(23)(34)(41)\;.
\end{equation}

Any other combination can be reduced to these by means of the 
harmonic cyclic identity 

\begin{equation}\label{hci}
  (12)(34)+(13)(42)+(14)(23)= 0
\end{equation}

following from the property of the $\epsilon^{ij}$ contraction. 
Secondly, given four space-time points, one can construct two 
independent conformal invariants,\footnote{Here is a simple 
explanation, very much in the spirit of the Q{\&}S frame argument 
above. The translations $P_\mu$ and the conformal boosts $K_\mu$ 
act on $x^\mu$ as linear and non-linear shifts, correspondingly. 
So, the combined action of both of them can define a special 
P{\&}K frame in which there are only two out of the four 
space-time variables $x^\mu_{1,2,3,4}$ left. Out of them we can 
make three Lorentz invariants (the two squares and the scalar 
product). Finally, dilation invariance requires that we take the 
two independent ratios of those.} the cross-ratios 

\begin{equation}\label{crrt}
 s = {x^2_{14}x^2_{23}\over x^2_{12}x^2_{34}}\;, \qquad 
t = {x^2_{13}x^2_{24}\over x^2_{12}x^2_{34}}\;. 
\end{equation}

Consequently, the most general form of the leading component of 
the correlator (\ref{4ptco}) consistent with H-analyticity and 
conformal covariance is 

\begin{equation}\label{l04}
 {(12)^2(34)^2\over x^4_{12}x^4_{34}}a(s,t) 
+{(14)^2(23)^2\over x^4_{14}x^4_{23}}b(s,t) +  
{(12)(23)(34)(41)\over x^2_{12}x^2_{23}x^2_{34}x^2_{41}}c(s,t)\;. 
\end{equation}

Here we see the three independent harmonic structures (\ref{3str}) 
completed to product of propagators. Such products already have 
the required conformal properties of the correlator, so the only 
freedom left are the three arbitrary coefficient functions $a,b,c$ 
of the invariant cross-ratios. Our aim will be to find constraints 
on these functions following from the full implementation of 
H-analyticity combined with superconformal covariance. 

The first step is to argue, just like in the three-point case, 
that there exists a special frame in superspace in which there are 
no $\theta^+$ left:  

\begin{equation}\label{QSfr4}
\mbox{Q{\&}S frame:}\qquad  
({\theta^{+\alpha,\dot\alpha}_{1}})'=({\theta^{+\alpha,\dot\alpha}_{2}})'= 
({\theta^{+\alpha,\dot\alpha}_{3}})'=({\theta^{+\alpha,\dot\alpha}_{4}})'=0 
\;. 
\end{equation}

As explained above, this can be achieved by fully exploiting the 
four spinor parameters contained in the doublets $\epsilon^i$ of Q 
and $\eta^i$ of S supersymmetry to shift away all four $\theta^+$. 
The existence of such a frame implies that the completion of the 
leading component (\ref{l04}) to a full superfield is always 
possible and is uniquely determined by Q and S supersymmetry. To 
obtain this completion one could, in principle, find the finite 
transformation to the frame (\ref{QSfr4}). However, unlike the 
case of the linear Q supersymmetry, S supersymmetry acts on the 
coordinates in a very non-linear way and the practical realisation 
of this step is not at all easy. Fortunately, as we shall explain 
below, for our purposes we shall only need to know the first 
non-trivial level in the $\theta^+$ expansion of the correlator. 

The above discussion makes it clear that no further constraints on 
the coefficient functions $a,b,c$ originate from conformal 
supersymmetry alone. This only takes place when we try to impose 
H-analyticity. There are two possible approaches in doing so. One 
is to use the form (\ref{hanG4}) of the constraint and try to 
solve it level by level in the $\theta^+$ expansion. The problem 
here is that this expansion is very complicated (assuming that we 
have already found it, which is in itself not an easy task). A 
much more efficient approach is to use the alternative form  

\begin{equation}\label{hanG2-}
  (D^{--}_4)^3 G^{(2,2,2,2)}(1\vert 2\vert 3\vert 4) = 0\;.
\end{equation} 

The advantage is that we can study this constraint in the Q{\&}S 
frame (\ref{QSfr4}) where there are no $\theta^+$'s. This results 
in substantial technical simplifications. The same trick is not 
allowed in the form (\ref{hanG4}) because of the harmonic 
singularities (see the discussion after eq. (\ref{propa})).

\subsubsection{An example of H-analyticity in the Q frame}

In order to better understand the idea of this approach, we are 
going to redo the derivation of the propagator (\ref{propa}), but 
this time starting form the alternative form of the H-analyticity 
condition 

\begin{equation}\label{altfo}
  (D^{--}_1)^2 G^{(1,1)}(1\vert 2) = 0\;.
\end{equation}

We begin by going to the Q frame (\ref{Qfr}). There the left-hand 
side of eq. (\ref{altfo}) does not depend on $\theta^+$ but can 
still depend on $\theta^-$. In particular, since the operator 
$D^{--}$ converts $\theta^+$ into $\theta^-$ (see (\ref{d--})), 
some terms in the $\theta^+$ expansion of $G^{(1,1)}$ may survive 
the transformation (\ref{epsii}), (\ref{epsi}). Therefore we 
should proceed in the following order. 

{\sl Step 1.} Expand $G^{(1,1)}$ in  $\theta_1^+$ up to the order  
$\theta_1^+\bar\theta_1^+$ (still in the old frame): 

\begin{equation}\label{expt1}
  G^{(1,1)} = g^{(1,1)}(x^2_{12},u_1,u_2) + \theta_1^{+\alpha} 
\bar\theta_1^{+\dot\alpha} 
\gamma^{(-1,1)}_{\alpha\dot\alpha}(x^2_{12},u_1,u_2) + \ldots 
\end{equation}

There is no need to keep terms containing $\theta_2^+$ or higher 
orders in $\theta_1^+$ because they cannot be ``rescued" by 
$(D_1^{--})^2$ and will vanish after the transformation to the Q 
frame. Indeed, the function $G^{(1,1)}$ carries no R weight, so 
the Grassmann variables have to appear in its expansion in pairs 
$\theta^+\bar\theta^+$. So, only the term 
$\theta_1^+\bar\theta_1^+\ \Rightarrow \ \theta_1^-\bar\theta_1^-$ 
can survive in the Q frame. 

{\sl Step 2.} Differentiate the expansion (\ref{expt1}) with 
$(D_1^{--})^2$ keeping only terms without any $\theta^+$. The 
expansion of $(D_1^{--})^2$ is 

\begin{equation}\label{D-2}
(D_1^{--})^2 = (\partial_1^{--})^2 
-4i\theta^-_1\dslash_1\bar\theta^-_1 \partial_1^{--} + 
(\theta^-_1\partial^-_1 + \bar\theta^-_1\bar\partial^-_1)^2 
-2(\theta^-_1)^2(\bar\theta^-_1)^2\square_1 \;. 
\end{equation}

We have dropped the terms linear in $2\theta^-_1\partial^-_1 + 
\bar\theta^-_1\bar\partial^-_1$ because they only convert one 
$\theta_1^+$ into $\theta_1^-$, and the remaining $\theta_1^+$ in 
the bilinear combination will vanish in the Q frame. When applied 
to (\ref{expt1}), this operator gives 

\begin{eqnarray}
(D_1^{--})^2G^{(1,1)} &=& (\partial_1^{--})^2g^{(1,1)} \nonumber\\ 
  && +\theta_1^{-\alpha} 
\bar\theta_1^{-\dot\alpha} 
\left[2\gamma^{(-1,1)}_{\alpha\dot\alpha} - 4i 
\partial_{1\;\alpha\dot\alpha}\partial_1^{--}g^{(1,1)} 
\right] \nonumber\\ && 
-2(\theta^-_1)^2(\bar\theta^-_1)^2\square_1g^{(1,1)} \nonumber\\ 
&& +\mbox{$\theta^+$ terms} \nonumber\\ &=& 0\;.\label{2pcons} 
\end{eqnarray}

{\sl Step 3.} Make the transformation to the Q frame. The 
left-hand side of eq. (\ref{2pcons}) is superconformally 
covariant, so it is multiplied by the weight factor 
$\Lambda(1)+\Lambda(2)$. Since it is supposed to vanish, this 
transformation just amounts to neglecting all the $\theta^+$ 
dependence (already taken into account at steps 1 and 2).     

The resulting constraint (\ref{2pcons}) involves three levels in 
its $\theta^-$ expansion. 

{\sl Level 0} or $(\theta^-\bar\theta^-)^0$: 

\begin{equation}
 (\partial_1^{--})^2g^{(1,1)}(x^2_{12},u_1,u_2) = 0 \quad 
\Rightarrow \quad g^{(1,1)} = (12)g(x^2_{12})\;. 
\end{equation}

This constraint uniquely fixes the harmonic dependence of the 
component $g^{(1,1)}$. The combination of harmonics $(12) \equiv 
u^{+i}_1 u^+_{2i}$ is the only one which is $SU(2)$ invariant, has 
the right charges and is annihilated by the lowering operator 
$(\partial_1^{--})^2$. 

{\sl Level 1} or $(\theta^-\bar\theta^-)^1$: 

\begin{equation}
  \gamma^{(-1,1)}_{\alpha\dot\alpha}(x^2_{12},u_1,u_2) = 2i 
\partial_{1\;\alpha\dot\alpha}\partial_1^{--}g^{(1,1)}(x^2_{12},u_1,u_2) =2i(12) 
\partial_{1\;\alpha\dot\alpha}\;g(x^2_{12})\;.  
\end{equation}
 
{\sl Level 2} or $(\theta^-\bar\theta^-)^2$: 

\begin{equation}\label{dropd}
  \square_1g^{(1,1)}(x^2_{12},u_1,u_2) =0 \quad 
\Rightarrow \quad \square_1 g(x^2_{12})=0 \quad \Rightarrow \quad 
g(x^2_{12}) = {C\over x^2_{12}} 
\end{equation}

where $C$ is an arbitrary constant. We recall that we are only 
interested in the two-point function away from the coincident 
point, so we can drop the delta-function $\delta(x_{12})$ in 
(\ref{dropd}). 

We should mention that in this example we have made no use of 
conformal invariance or S supersymmetry. Actually, this two-point 
function is in a sense overdetermined. We have already seen that 
by imposing H-analyticity just at level 0 and then invoking 
dilation covariance (part of the conformal symmetry), we arrived 
at the same result. This, however, is an exceptional property of 
the propagator (the charges $+1$ two-point function). The typical 
situation is illustrated by the charges $+2$ two-point function  
(\ref{crrr}). It is not hard to show that it remains H-analytic to 
all orders in the $\theta$ expansion even if we replace the 
denominator by any function of $\hat x^2_{12}$. So,  its form 
cannot be determined without some extra input (dilation covariance 
in this case). The explanation of this fact can be traced back to 
the different implications of H-analyticity for superfields of 
charges $+1$ and $+2$: for the former it is an on-shell condition 
and for the latter an off-shell one (see (\ref{Hanq}), 
(\ref{linm})). 

Let us summarise the above example. Using the Q frame we have been 
able to solve the H-analyticity constraint (\ref{altfo}) to all 
relevant orders in the $\theta$ expansion. In the process we only 
used the $\theta^+$ expansion of the two-point function 
$G^{(1,1)}$ to the first non-trivial order (level 1) (see 
(\ref{expt1})). At no point we encountered delta-type harmonic 
singularities which cannot coexist with the singular nature of the 
transformation to the Q frame. On the contrary, starting with the 
form (\ref{hanG}), we would have to find out the $\theta^+$ 
expansion of $G^{(1,1)}$ to all orders (in this case it can go up 
to level 4) and then solve the H-analyticity constraint order by 
order. The technical advantages of the use of the alternative form 
of the H-analyticity constraint and of the Q (or Q{\&}S) frame 
result in major simplifications in the case of the four-point 
function.

\subsubsection{Superconformal covariance at level 1}

Now we come back to the four-point function (\ref{4ptco}). Eq. 
(\ref{l04}) represents the solution to the H-analyticity 
constraint and to the conformal covariance condition at level 0. 
We have also argued that the possibility to go to the Q{\&}S frame 
(\ref{QSfr4}) guarantees the existence of a unique completion of 
this level 0 component to a full superfield. The way to find this 
completion consists of two steps. The first is to put hats on all 
the $x$'s in the denominators, thus reconstructing the products of 
full propagators. We already know that such products have the 
required superconformal properties of the correlator. The second 
step is to complete the conformal cross-ratios $s$ and $t$ in the 
coefficient functions $a,b,c$ to full superconformal invariants 
$\hat s$ and $\hat t$. Then the full correlator consistent with 
superconformal symmetry will have the form 

\begin{equation}\label{l04hat}
G^{(2,2,2,2)}(1\vert 2\vert 3\vert 4)= {(12)^2(34)^2\over \hat 
x^4_{12}\hat x^4_{34}}a(\hat s,\hat t) +{(14)^2(23)^2\over \hat 
x^4_{14}\hat x^4_{23}}b(\hat s,\hat t) +  {(12)(23)(34)(41)\over 
\hat x^2_{12}\hat x^2_{23}\hat x^2_{34}\hat x^2_{41}}c(\hat s,\hat 
t)\;. 
\end{equation}

To find $\hat s$ and $\hat t$ to all orders in the four 
$\theta^+$'s is a very non-trivial task (the expansion goes up to 
level 8, although Q supersymmetry helps bring it down to level 4). 
Fortunately, the example above has taught us that we only need one 
level 1 term. So, we are looking for $\hat s$ in the form 

\begin{equation}\label{hats}
  \hat s = s + \sum^4_{a,b=1}\theta^{+\alpha}_a
S_{ab\;\alpha\dot\alpha}\bar\theta^{+\dot\alpha}_b + 
O((\theta^+\bar\theta^+)^2)\;. 
\end{equation}

What we really need is just the coefficient 
$S_{44\;\alpha\dot\alpha}$. Indeed, although the three derivatives 
$(D_4^{--})^3$ in (\ref{hanG2-}) can convert a maximum of three 
$\theta_4^+$ into $\theta_4^-$, only the term 
$\theta_4^+\bar\theta_4^+\ \Rightarrow \ \theta_4^-\bar\theta_4^-$ 
can survive in the Q{\&}S frame. 

The coefficient $S_{44\;\alpha\dot\alpha}$ can be solved for from 
a set of linear equations. It is obtained by performing a combined 
Q (\ref{Qsusy}) and S (\ref{7.9.1c}) supersymmetry transformation 
on $\hat s$ and demanding that it be invariant. In doing so we 
shall only keep the terms linear in $\bar\theta^+_4$ since only 
they involve the coefficients $S_{a4}$: 

\begin{equation}\label{varsh}
 \delta_{Q+S}\hat s=0\ \Rightarrow\ \left[4is(\epsilon^{-\alpha}_4
 + x^{\alpha\dot\beta}_4 \bar\eta^-_{4\dot\beta})\left({x_{14}\over x^2_{14}} 
- {x_{34}\over x^2_{34}}\right)_{\alpha\dot\alpha} + 
\sum^4_{a=1}(\epsilon^{+\alpha}_a 
 + x^{\alpha\dot\beta}_a \bar\eta^+_{a\dot\beta})S_{a4\;\alpha\dot\alpha} 
\right]\bar\theta^{+\dot\alpha}_4 = 0\;. 
\end{equation}

Here $(\epsilon,\eta)^\pm_a \equiv u^\pm_{ai}(\epsilon,\eta)^i$. 
Removing $\bar\theta^{+\dot\alpha}_4$ and the independent 
parameters $\epsilon,\eta$ from (\ref{varsh}), we obtain four 
linear equations (one for each harmonic projection of the two 
parameters) for the four coefficients $S_{a4}$: 

\begin{eqnarray}
  &&4isX_2 - \sum^3_{a=1}(a4)S_{a4} = 0\;, \quad
  \sum^4_{a=1}(4^-a)S_{a4} = 0\;, \nonumber\\
  && (12)x_{24}S_{24} + (13)x_{34}S_{34} = 0\;, \quad
(21)x_{14}S_{14} + (23)x_{34}S_{34} = 0\;. \label{system} 
\end{eqnarray}

Here and in what follows we use the vectors 

\begin{equation}
X_1 = {x_{14}\over x^2_{14}} - {x_{24}\over x^2_{24}}\;, \quad   
X_2 = {x_{14}\over x^2_{14}} - {x_{34}\over x^2_{34}}\;, \quad   
X_3 = {x_{34}\over x^2_{34}} - {x_{24}\over x^2_{24}} 
\end{equation}

having the useful properties 

\begin{equation}\label{Xide}
  {X^2_1\over X^2_3} ={1\over s}\;,\quad {X^2_2\over X^2_3} = {t\over s}\;, 
\quad {2X_1\cdot X_3\over X^2_3} ={1+s-t\over s}\;, \quad 
{2X_2\cdot X_3\over X^2_3} = {1-s-t\over s}\;. 
\end{equation}

Solving eqs. (\ref{system}) is a straightforward calculation. The 
result for the coefficient $S_{44}$ is 

\begin{eqnarray}
 {(12)^2(34)^2Y\over 4i}S_{44} &=&  - (12)(13)(23)tX_3\nonumber\\
  &-&[(12)(13)(23)s + (12)^2(34)(34^-)s\nonumber\\
&+& (12)(23)(34)(14^-)(1+s-t) + (14)(14^-)(23)^2]X_2  \label{S44} 
\end{eqnarray}

where 

\begin{equation}\label{Y}
  Y = 1 + {1+s-t\over s}U +{1\over s}U^2\;, \qquad U = {(14)(23)\over 
(12)(34)}\;. 
\end{equation}

Note that the vector (\ref{S44}) can be rewritten in the 
equivalent form 

\begin{equation}\label{idS44}
{(12)^2(34)^2Y\over 4i}S_{44} = -{s\over 
2}\partial^{--}_4[(12)^2(34)^2Y] X_2 + (12)(13)(23)[-tX_3 + 
{1\over 2}(1-s-t)X_2]\;.  
\end{equation}
 
In in a similar manner we can find the relevant term in the 
expansion of $\hat t$ 

\begin{equation}\label{hatt}
  \hat t = t + \sum^4_{a,b=1}\theta^{+\alpha}_a
T_{ab\;\alpha\dot\alpha}\bar\theta^{+\dot\alpha}_b + \ldots 
\end{equation} 

The result for $T_{44}$ can be easily obtained from (\ref{S44}) by 
exchanging $1\ \leftrightarrow\ 2$ which implies $s\ 
\leftrightarrow\ t$, $X_1\ \rightarrow\ -X_1$, $X_2\ \rightarrow\ 
-X_3$, $Y\ \rightarrow\ s/t\; Y$. 

The expansion of $\hat s$, $\hat t$ is not the only source of 
$\theta^+_4\bar\theta^+_4$ terms. Another contribution comes from 
expanding $\hat x^2_{a4}$, $a=1,2,3$ in the propagators in 
(\ref{l04hat}): 

\begin{equation}\label{hatcontr}
  \hat x^2_{a4} = x^2_{a4} -4i{(a4^-)\over 
(a4)}\theta^+_4x_{a4}\bar\theta^+_4 + \ldots 
\end{equation}

(recall (\ref{hataa})). Collecting all of these contributions, we 
can write down 

\begin{eqnarray}
 G^{(2,2,2,2)}(1\vert 2\vert 3\vert 4) &\equiv& 
\Pi(\hat x^2_{ab})\cdot f(\hat s,\hat t) \nonumber\\ 
  &=& G^{(2,2,2,2)}\vert_{\mbox{\scriptsize level 0}} \label{exp4p}\\
&+& \theta^+_4\left[-4i\sum^3_{a=1}{(a4^-)\over (a4)}x_{a4} 
{\partial\Pi\over\partial x^2_{a4}}\cdot f + \Pi\cdot (f_s S_{44} 
+ f_t T_{44}) \right]\bar\theta^+_4 + \ldots \nonumber 
\end{eqnarray}

Here $\Pi\cdot f$ is a shorthand for the product of propagators 
($\Pi$) and coefficient functions ($f$) in (\ref{l04hat}) and 
$f_{s,t}= \partial f/\partial(s,t)$. This accomplishes Step 1 of 
our programme for imposing H-analyticity in the form 
(\ref{hanG2-}).

\subsubsection{Constraints from H-analyticity}

Step 2 consists of differentiating the expression (\ref{exp4p}) 
with the operator $(D^{--}_4)^3$ 

\begin{eqnarray}
 (D_1^{--})^3 &=& (\partial_1^{--})^3\nonumber\\
  &+& 3\partial_1^{--}(\theta^-_1\partial^-_1 + 
\bar\theta^-_1\bar\partial^-_1)^2 - 
6i(\partial_1^{--})^2\theta^-_1\dslash_1\bar\theta^-_1\nonumber\\ 
&-& 6i\theta^-_1\dslash_1\bar\theta^-_1(\theta^-_1\partial^-_1 + 
\bar\theta^-_1\bar\partial^-_1)^2 - 
6\partial_1^{--}(\theta^-_1)^2(\bar\theta^-_1)^2\square_1\nonumber\\ 
&+& \mbox{irrelevant terms}\;.   \label{D-2} 
\end{eqnarray}

The irrelevant terms in (\ref{D-2}) are those which convert an odd 
number of $\theta_1^+$ into $\theta_1^-$ and thus disappear in the 
Q frame. We then apply (\ref{D-2}) to (\ref{exp4p}) and collect 
all the terms at levels 0, 1 and 2. In fact, we have already 
solved the  H-analyticity constraint at level 0 in (\ref{l04}). 
So, it remains to examine the constraints at levels 1 and 2. It is 
not hard to see that they take the form 

\begin{eqnarray}
  &&\mbox{Level 1:}\qquad \partial^{--}_4  A^{\mu} = 0\;,  \label{4lev1}\\
  &&\mbox{Level 2:}\qquad \partial_{4\mu}  A^{\mu} = 0  \label{4lev2}
\end{eqnarray}

where 

\begin{equation}\label{vector}
  A^{\mu} = -i\partial^{--}_4\partial^\mu_4\left(\Pi\cdot f\right)
-4i\sum^3_{a=1}{(a4^-)\over (a4)}x^\mu_{a4} 
{\partial\Pi\over\partial x^2_{a4}}\cdot f  + \Pi\cdot (f_s 
S^\mu_{44} + f_t T^\mu_{44})\;. 
\end{equation}

After some simple algebra and using the relations 

\begin{equation}\label{derst}
  \partial^\mu_4 s = -2sX^\mu_2\;, \qquad \partial^\mu_4 t = 2tX^\mu_3\;, 
\end{equation}

we obtain 

\begin{equation}\label{vectB}
 A^{\mu} = {2ic\over s}\;{(12)(13)(23)\over x^4_{12}x^4_{34}}\; X^\mu_2
+[2is X^\mu_2\partial^{--}_4\Pi + S^\mu_{44}\Pi]\cdot f_s + [-2it 
X^\mu_3\partial^{--}_4\Pi + T^\mu_{44}\Pi]\cdot f_t\;. 
\end{equation}

The terms in the brackets are computed with the help of 
(\ref{idS44}), e.g. 

\begin{eqnarray}
  [2is X_2\partial^{--}_4\Pi + S_{44}\Pi]\cdot 
f_s &=& 2is X_2 (12)^2(34)^2Y\; \partial^{--}_4\left({\Pi\cdot 
f_s\over (12)^2(34)^2Y} \right) \label{halfofit}\\ 
  &+&4i(12)(13)(23)[-tX_3 + {1\over 
2}(1-s-t)X_2]\; {\Pi\cdot f_s\over (12)^2(34)^2Y} \;.\nonumber  
\end{eqnarray}

Further, it is convenient to use the harmonic cross-ratio $U$ 
introduced in (\ref{Y}) and rewrite 

\begin{equation}
  {\Pi\cdot f_{s}\over (12)^2(34)^2Y} = {1\over 
x^4_{12}x^4_{34}}\ {a_{s} - {c_{s}\over s}U + {b_{s}\over 
s^2}U^2\over 1 + {1+s-t\over s}U +{1\over s}U^2}\;, 
\end{equation}

after which the harmonic derivative $\partial^{--}_4$ in 
(\ref{halfofit}) can be computed using the identity 
$$
\partial^{--}_4 U = - {(12)(13)(23)\over (12)^2(34)^2}\;.
$$
Repeating the same procedure for the other bracket in 
(\ref{vectB}) and collecting all the terms proportional to the 
vector $X_2$, we obtain the following contribution to the vector 
$A^\mu$: 

\begin{equation}\label{ready}
2i{(12)(13)(23)\over x^4_{12}x^4_{34}}\;\left[{c\over s} + \beta_0 
{1 + {\beta_1\over \beta_0} U +{\beta_2\over \beta_0} U^2 \over  1 
+ {1+s-t\over s}U +{1\over s}U^2}\right] \;  X^\mu_2 
\end{equation}

where 

\begin{eqnarray}
 \beta_0 &=& c_s + 2(1-t)a_s -2ta_t\;, \nonumber\\
 \beta_1 &=&  2a_s - {2\over 
s}b_s + {2t\over s}c_t + {s+t-1\over s}c_s \;, \\ 
 \beta_2 &=& -{2\over s}b_s -{2t\over s^2}b_t -{1\over s}c_s\;. \nonumber 
\end{eqnarray}
 
We still have to compute the $X_3$ contribution to $A^\mu$, but 
even before this we can already impose the level 1 constraint 
(\ref{4lev1}) on the $X_2$ contribution (the vectors $X_2$ and 
$X_3$ are linearly independent and $\partial^{--}_4 X_2 = 
\partial^{--}_4 X_3 = 0$). The first term in (\ref{ready}) does 
not depend on $u_4$. The second term is the ratio of two 
polynomials of degree 2 in the cross-ratio $U$. It is easy to see 
that its derivative vanishes only if the two polynomials are 
equal, 

\begin{equation}
  1 + {\beta_1\over \beta_0} U 
+{\beta_2\over \beta_0} U^2 = 1 + {1+s-t\over s}U +{1\over 
s}U^2\;. 
\end{equation}

Comparing the coefficients in front of $U$ and $U^2$, we obtain 
the following constraints: 

\begin{eqnarray}
  &&c_s = (t-1)a_s + ta_t -b_s - {t\over s}b_t\;, \nonumber\\
  &&c_t = -sa_s -sa_t +b_s + {t-1\over s}b_t \label{cts}
\end{eqnarray}

constituting our main result. This is a set of two linear 
first-order partial differential equations for the three 
coefficient functions $a,b,c$. These equations can only determine 
two of the three functions. Indeed, let us make the change of 
variables 

\begin{equation}\label{chvar}
  a = \alpha +\gamma +{1\over s}b\;, \qquad c = -{s}\gamma + {t-s-1\over s}b\;, 
\end{equation}

after which (\ref{cts}) becomes 

\begin{eqnarray}
  &&\gamma_s = -\alpha_s - \alpha_t\;, \nonumber\\
  &&t\gamma_t +\c = s\alpha_s + (s-1)\alpha_t\label{ctss}
\end{eqnarray}

and we see that $b$ has dropped out. The set of first-order 
coupled differential equations (\ref{ctss}) can be equivalently 
rewritten as a set of second-order independent ones: 

\begin{eqnarray}
s\alpha_{ss} + t\alpha_{tt} + (s+t-1)\alpha_{st} + 2(\alpha_s + 
\alpha_t) &=& 0\;, \nonumber\\ s\gamma_{ss} + t\gamma_{tt} + 
(s+t-1)\gamma_{st} + 2(\c_s + \c_t) &=& 0 \; . \label{2nd} 
\end{eqnarray}

These constraints are the same as eqs. (\ref{sameres}) obtained in 
section 2 in the coordinate approach. 

To compute the contribution to $A^\mu$ proportional to  $X_3$ we 
go through the same steps. This time we do not find any new 
constraints. The final form of the vector $A^\mu$ is 

\begin{equation}\label{finalA}
  A^\mu = 2i(12)(13)(23)\left(A{X^\mu_2\over 
x^4_{12}x^4_{34}s} + B{X^\mu_3\over x^4_{12}x^4_{34}t}\right) 
\end{equation}

where 

\begin{equation}\label{AB}
 A = sc_s +2(1-t)sa_s -2sta_t+c\;, \qquad B = -{t^2\over s}(c_t 
+2sa_s+2sa_t)\;. 
\end{equation}
 
The remaining step is to impose the level 2 constraint  
(\ref{4lev2}). This is facilitated by the useful property 
$$
\partial_{4\mu}\left({X^\mu_2\over 
x^4_{12}x^4_{34}s} \right) = \partial_{4\mu}\left({X^\mu_3\over 
x^4_{12}x^4_{34}t} \right) = 0 
$$  
of the basis vectors in (\ref{finalA}), so we only have to 
differentiate the scalar coefficients $A$ and $B$. Using the 
identities (\ref{derst}) and (\ref{Xide}), we obtain the 
constraint 

\begin{equation}
\partial_{4\mu} A^\mu=0 \quad \Rightarrow \quad -{t\over s}A_s - B_t 
+{1-s-t\over 2s} \left({t\over s}A_t + {s\over t}B_s \right) =0\;. 
\end{equation}

Making the change of variables (\ref{chvar}) and after some 
algebra we discover that this is a corollary of the second-order 
differential equations (\ref{2nd}). So, level 2 does not give rise 
to any new constraints. 


\section{Conclusions}

We see that using either method of analysing analyticity for the 
four-point charge 2 correlator leads to the same result: The 
requirements of H-analyticity and superconformal covariance yield 
constraints which fix the form of the four-point correlator 
(\ref{4ptco}) up to an arbitrary function of the conformal 
cross-ratios. An interesting solution to these constraints is 
provided by the explicit computation of the correlator at two 
loops carried out in \cite{4pt',ehssw2}. The result for the level 
0 component is 

\begin{equation}\label{explf}
 \Phi(s,t)\left[{(12)^2(34)^2\over x^4_{12}x^4_{34}} 
+{(14)^2(23)^2\over x^4_{14}x^4_{23}}s + {(12)(23)(34)(41)\over 
x^2_{12}x^2_{23}x^2_{34}x^2_{41}}(t-s-1) \right]\;. 
\end{equation}

Here $\Phi(s,t)= \Phi(t,s) = {1\over s}\Phi({1\over s},{t\over 
s})$ is a function given by the one-loop scalar box integral. This 
solution is symmetric under the exchange $1\;\leftrightarrow\;3$ 
and is determined by the asymptotic behaviour 
$\lim_{x_{14}\rightarrow0} c(s,t) =0$ (see \cite{4pt'} for 
details). It is interesting to note how the result (\ref{explf}) 
was obtained: the two-loop calculation in  \cite{4pt'} only 
provided us with the explicit form of the first two terms in 
(\ref{explf}); the third term was given as a complicated two-loop 
integral; the subsequent use of the differential equations  
(\ref{2nd}) in conjunction with the boundary conditions following 
from the known asymptotic behaviour of the two-loop integral 
allowed us \cite{ehssw2}} to solve for the third term as in  
(\ref{explf}).  

We have shown for a number of cases that harmonic analyticity at 
all orders in an expansion of a correlator in the odd variables is 
assured by the constraints arising from the lowest and the linear 
order. We conjecture that this is a general feature.

\vspace{20pt} {\bf Acknowledgements:} This work was supported in 
part by the British-French scientific programme Alliance (project 
98074), by the EU network on Integrability, non-perturbative 
effects and symmetry in quantum field theory (FMRX-CT96-0012) and 
by the grant INTAS-96-0308.


\begin{thebibliography}{99}

\bibitem{maldacena} 
J. Maldacena, {\sl The large N limit of superconformal field 
theories and supergravity}, Adv. Theor. Math. Phys. {\bf 2} (1998) 
231-252, hep-th/9711200; S.S. Gubser, I.R. Klebanov and A.M. 
Polyakov, {\sl Gauge theory correlators from noncritical String 
theory}, Phys. Lett. {\bf B428} (1998) 105, hep-th/9802109; E. Witten, 
{\sl Anti-de  Sitter space and holography}, Adv. Theor. Math. 
Phys. {\bf 2} (1998) 253-291, hep-th/9802150. 

\bm{ads} Hong-Liu and A.A. Tseytlin, {\sl On four-point functions 
in the CFT/AdS correspondence}, hep-th/9807097; D.Z. Freedman, S. 
Mathur, A. Matusis and L. Rastelli, {\sl Comments on four-points 
functions in the CFT/AdS correspondence}, hep-th/9808006; E. d' 
Hoker and D.Z. Freedman, {\sl Gauge boson exchange in AdS(d+1)}, 
hep-th/9809179; E. D'Hoker, D.Z. Freedman, S.D. Mathur, A. 
Matusis, L. Rastelli, {\sl Graviton exchange and complete four 
point functions in the AdS / CFT correspondence} hep-th/9903196; 
E. D'Hoker, D.Z. Freedman, L. Rastelli, {\sl AdS / CFT four point 
functions: How to succeed at z integrals without really trying} 
hep-th/9905049.

\bm{4pt} M. Bianchi, M. Green , S. Kovacs and G. Rossi, {\sl 
Instantons in supersymmetric Yang-Mills and D instantons in IIB 
superstring theory}, JHEP {\bf 08} (1998) 13, hep-th/9807033; N. 
Dorey, T. Hollowood, V. Khoze, M. Mattis and S. Vandoren, {\sl 
Mulitinstantons and Maldacena's conjecture} hep-th/9811060; 
F. Gonzalez-Rey, I. Park and K. Schalm, {\sl A note on four-point 
functions of conformal operators in $N=4$ super-Yang Mills}, Phys. 
Lett. {\bf B448} (1999) 37-40, hep-th/9811155;

\bm{4pt'}  B. Eden, P.S. Howe, C. Schubert, E. Sokatchev and P.C. 
West, {\sl Four-point functions in N=4 supersymmetric Yang-Mills 
theory at two loops}, Nucl. Phys. {\bf B557} (1999) 355-379, 
hep-th/9811172.

\bibitem{hw1}
P.S. Howe and P.C. West, {\sl Non-perturbative Green's functions 
in theories with extended superconformal symmetry}, 
hep-th/9509140; {\sl OPEs in four-dimensional superconformal field 
theories}, Phys. Lett. {\bf B389} (1996) 273-279, hep-th/9607060; 
{\sl Is $N=4$ Yang-Mills soluble?} (1996) in Moscow 1996 Physics, 
622-626, hep-th/9611074. 

\bibitem{hw2} P.S. Howe and P.C. West, {\sl Superconformal invariants and
extended supersymmetry}, Phys. Lett. {\bf B400} (1997) 307-313, 
hep-th/9611075. 

\bm{op} B. Conlong and P. West, {\sl Anomalous dimensions of 
fields in a supersymmetric quantum field theory at a 
renormalization group fixed point}, J. Phys. {\bf A26} (1993) 
3325; H. Osborn, {\sl N=1 superconformal symmetry in 
four-dimensional quantum field theory} Ann. Phys. (N.Y.) {\bf 272} 
(1999) 243-294, hep-th/980804; J-H. Park, {\sl Superconformal 
symmetry and correlation functions} hep-th/9903230. 

\bibitem{hsw} P.S. Howe, E. Sokatchev and P.C. West,
{\sl Three-point functions in N=4 Yang-Mills}, Phys. Lett. {\bf 
B444} (1998) 341-351, hep-th/9808162. 

\bm{3pt} S. Lee, S. Minwalla, M. Rangamani and N. Seiberg, {\sl 
Three-point functions of chiral operators in D=4, N=4 SYM at large 
N}, Adv. Theor. Math. Phys {\bf 2} (1998) 697-718, hep-th/9806074; 
F. Gonzalez-Rey, B. Kulik and I.Y. Park, {\sl Non-renormalisation of 
two and three point correlators of N=4 SYM in N=1 superspace}, 
Phys. Lett. {\bf B455} (1999) 164-170, hep-th/9903094;
S. Penati, A. Santambrogio and D. Zanon, {\sl Two-point functions
of chiral operators in $N=4$ SYM at order $g^4$}, Bicocca-FT-99-30,
hep-th/9910197.

\bm{dhfs} D. Z. Freedman, S.D. Mathur, A. Matusis and L. Rastelli, 
{\sl Correlation functions in the CFT(d) / AdS(d+1) 
correspondence}, Nucl. Phys. {\bf B546} (1999) 96-118, hep-th/9804058. 

\bm{dhfs2} E. D'Hoker, D.Z. Freedman and W. Skiba, {\sl Field 
theory tests for correlators in the AdS/CFT correspondence}, Phys. 
Rev. {\bf D59} (1999) 45008, hep-th/9807098. 
 
\bibitem{ken1} K. Intriligator, {\sl Bonus symmetries of N=4
super-Yang-Mills correlation functions via AdS duality}, Nucl. 
Phys. {\bf B551} (1999) 575-600, hep-th/9811047. 

\bm{ehssw2} B. Eden, P.S. Howe, C. Schubert, E. Sokatchev and P.C. 
West, {\sl Simplifications of four-point functions in N=4 
supersymmetric Yang-Mills theory at two loops}, Phys. 
Lett. {\bf B466} (1999) 20-26, hep-th/9906051. 

\bm{ehssw3} B. Eden, P.S. Howe, C. Schubert, E. Sokatchev and P.C. 
West, {\sl Extremal correlators in four-dimensional SCFT}, Phys. 
Lett. {\bf B472} (2000) 323-331, hep-th/9910150.

\bibitem{dfmmrext}
E. D'Hoker, D.Z. Freedman, S. D. Mathur, A. Matusis and L. 
Rastelli, {\sl Extremal correlators in the AdS/CFT 
correspondence}, MIT-CTP-2893, hep-th/9908160. 

\bibitem{arufrov}
G. Arutyunov and S. Frolov, {\sl Some cubic couplings in type IIB
supergravity on $AdS_5 \times S^5$ and three-point functions in $SYM_4$
at large $N$}, LMU-TPW 99-12, hep-th/9907085; {\sl Scalar quartic
couplings in type IIB supergravity on $AdS_5 \times S^5$}, LMU-TPW 99-23,
hep-th/9912210.

\bibitem{biakov}
M. Bianchi and S. Kovacs, {\sl Nonrenormalization of extremal 
correlators in $N=4$ SYM theory}, ROM2F-99-31, hep-th/9910016. 

\bm{gikos} A. Galperin, E. Ivanov, S. Kalitzin, V. Ogievetsky and 
E. Sokatchev, {\sl Unconstrained $N=2$ matter, Yang-Mills and 
supergravity theories in harmonic superspace}, Class. Quant. Grav. 
{\bf 1} (1984) 469. 

\bibitem{hh} G.G. Hartwell and P.S. Howe {\sl (N, p, q) harmonic
superspace} Int. J. Mod. Phys. {\bf A10} (1995) 3901-3920, 
hep-th/9412147; {\sl A superspace survey}, Class. Quant. Grav. 
{\bf 12} (1995) 1823-1880. 

\bibitem{af} L. Andrianopoli and S. Ferrara,
{\sl K-K excitations on $AdS_5\times S^5$ as $N=4$ ``primary'' 
superfields}, Phys. Lett. {\bf B430} (1998) 248-253, 
hep-th/9803171. 

\bibitem{ehw} B. Eden, P.S. Howe and P.C. West,
{\sl Nilpotent invariants in N=4 SYM}, Phys. Lett. {\bf B463} (1999)
19-26, hep-th/9905085. 

\bibitem{hssw}
P.S. Howe, C. Schubert, E. Sokatchev and P.C. West, {\sl Explicit 
construction of nilpotent covariants in $N=4$ SYM}, KCL-MTH-99-41, 
hep-th/9910011. 

\bibitem{petske}
A. Petkou and K. Skenderis, {\sl A non-renormalisation theorem for 
conformal anomalies}, hep-th/9906030. 

\bibitem{hsgr}
A. Galperin, E. Ivanov, V. Ogievetsky and E. Sokatchev, {\sl 
Harmonic supergraphs: Green functions}, Class. Quant. Grav. {\bf 
2} (1985) 601-616; {\sl Harmonic supergraphs: Feynman rules and 
examples}, Class. Quant. Grav. {\bf 2} (1985) 617-630. 

\bibitem{fradkin}
A. Galperin, E. Ivanov, V. Ogievetsky and E. Sokatchev, {\sl Conformal 
invariance in harmonic superspace}, preprint JINR E2-85-363 (1985) 
published in ``Quantum Field theory and Quantum Statistics", 
vol.2, 233-248, A.Hilger, Bristol (1987).

\bibitem{ken2}
K. Intriligator and W. Skiba, {\sl 
Bonus symmetry and the operator product expansion of N=4 
Super-Yang-Mills}, hep-th/9905020. 

\bm{nogo} P.S. Howe, K.S. Stelle and P.C. West, {\sl N=1 d=6 
harmonic superspace}, Class. Quant. Grav. {\bf 2} (1985), 815-821; 
 K.S. Stelle, {\sl Manifest realizations of extended 
supersymmetry}, Santa Barbara preprint NSF-ITP-85-001. 

\bm{book} A. Galperin, E. Ivanov, V. Ogievetsky and E. Sokatchev, 
{\sl Harmonic superspace}, CUP, to appear. 

\end{thebibliography}
\end{document}